\documentclass[preprint2]{aastex}

\begin{document}

\title{\bf Starburst or Seyfert? Adding a radio and far-infrared
perspective to the 
investigation of activity in composite galaxies}

\author{Tanya~L.~Hill\footnote{Present address: Melbourne Planetarium,
2 Booker St, Spotswood, VIC, 3015, Australia; email:thill@museum.vic.gov.au}}
\affil{School of Physics, University of Sydney}
\authoraddr{School of Physics, University of Sydney, NSW, 2006, Australia}
\author{Charlene A. Heisler\footnote{This paper is dedicated to
the memory of Charlene Heisler (deceased, 28 October 1999).}}
\affil{Research School for Astronomy and Astrophysics, Australian
National University}
\authoraddr{Research School for Astronomy and Astrophysics, Institute of 
Advanced Studies, Mount Stromlo and Siding Spring Observatories, The Australian National University, Private Bag,\\ Weston Creek
Post Office, Canberra, ACT, 2611, Australia\\}
\author{Ray P. Norris and John E. Reynolds}
\affil{Australia Telescope National Facility}
\authoraddr{ATNF, PO Box 76, Epping, NSW, 2121, Australia}
\and
\author{Richard W. Hunstead}
\affil{School of Physics, University of Sydney}
\authoraddr{School of Physics, University of Sydney, NSW, 2006, Australia}

\begin{abstract}

It was once common to regard Seyfert and starburst galaxies as
completely different types of object, but there is growing recognition
that these classifications refer to the extremes of a continuous
spectrum of galaxy types. In a previous study we investigated a sample
of galaxies with ambiguous optical emission-line ratios and concluded
from near-infrared spectroscopic observations that the sample
consisted of composite galaxies, containing both a starburst and an
active galactic nucleus (AGN). We now extend our study using radio 
synthesis and
long-baseline interferometer observations made with the Australia
Telescope, together with far-infrared IRAS observations, to discuss
the relative contribution of starburst and AGN components to the
overall luminosity of the composite galaxies. We find that only a
small fraction of the radio emission ($<$~10\%) can be attributed to
an AGN, and that the majority of the far-infrared emission ($>$~90\%)
is probably due to the starburst component. We also show that an AGN
contribution to the optical emission of as little as 10\% is
sufficient to account for the ambiguous line-ratio diagnostics.

\end{abstract}
 

\newpage

\section{Introduction}

The possible existence of a relationship between massive star 
formation and an active galactic nucleus (AGN) is a long standing,
controversial issue. An increasing number of galaxies have been shown
to contain both intense star formation, typical of a starburst galaxy,
as well as a non-thermal ionising source, attributed to an AGN.  Well
known examples of such composite galaxies are NGC 1068 (Thatte {\it
et~al.}\ 1997), NGC 7469 (Genzel {\it et~al.}\ 1995), Mrk 477 (Heckman
{\it et~al.}\ 1997) and Circinus (Oliva {\it et~al.} 1995), all of which
contain, in addition to a Seyfert nucleus, circumnuclear star
formation that contributes significantly to the total luminosity.  In
many such cases, the composite nature of the galaxies is based on
optical and NIR observations. However, radio observations can also be
a powerful tool for distinguishing starburst and AGN activity.

In general, galaxies dominated by starbursts exhibit diffuse radio
emission as the result of synchrotron radiation associated with
supernova remnants and cosmic rays (Sramek \& Weedman 1986).  Seyfert
galaxies, on the other hand, generally exhibit more compact emission
or, in the more radio-energetic cases, linear structures, identifiable
as jets and radio lobes (Wilson {\it et~al.}\ 1991).  The use of radio
morphology to distinguish the dominant activity within galaxies is
well established (Norris {\it et~al.} 1988a; Wilson 1988; Crawford {\it et~al.} 
1996).  As one example, NGC~1068, a known composite galaxy, has been
observed in the radio to contain both diffuse emission, presumably as
a result of star formation, and linear structure, attributed to a jet
(Wilson \& Ulvestad 1982; Wynn-Williams {\it et~al.} 1985; Gallimore 
{\it et~al.} 1997).

As a step towards exploring the possible relationship between
starburst and Seyfert activity within a homogeneous data
set, we have formed a sample of galaxies that are composite in nature.
The galaxies were chosen on the basis of optical spectroscopy and
investigated further using near-infrared (NIR) spectroscopy 
(Hill {\it et~al.}\ 1999: Paper 1). In this previous study, 
we found only one galaxy, from a sample of 12, to 
be dominated in the optical/NIR by star formation, with
minimal, if any, contribution from an AGN. However, the remaining
11 galaxies show evidence that they are composed of a
starburst population and dust heated to temperatures of
up to 1000~K by an AGN. In this paper, we present a radio
study of the sample in an attempt to place tighter limits on the 
AGN contribution.

The Australia Telescope Compact Array (ATCA) was used to image our
sample at 3 and 6~cm. These data are combined with additional
information such as radio spectral indices and far-infrared (FIR)
colours to investigate further the nature of the powering source.  We
also present long-baseline interferometry observations using the
Parkes-Tidbinbilla Interferometer (PTI; Norris {\it et~al.} 1988b). PTI is
sensitive to structures with sizes $\leq 0.1^{\prime\prime}$ (corresponding to
40~pc at the mean distance of the sample) at 13~cm and brightness
temperatures $>$10$^5$ K. PTI will therefore respond to the
non-thermal emission from compact cores of Seyferts with brightness
temperatures $\sim$10$^8$ K, yet is essentially blind to extended H~II
regions or supernovae associated with starbursts, which have typical
brightness temperatures of 10$^4$ K. However, recent studies suggest
that extremely luminous radio supernovae (RSNe) associated with
starbursts may also account for some of the emission detected with PTI
(Kewley {\it et~al.} 2000a; Smith {\it et~al.} 1998a).

The sample selection criteria are presented in Section~2. Details of
the observations and data reduction for the ATCA and PTI are given in
Section~3. Our results are presented and discussed in Section~4, with 
our conclusions in Section~5.  Throughout this paper, we adopt H$_0$ =
75 km s$^{-1}$ Mpc$^{-1}$ and q$_0$ = 0.5.

\section{Sample Selection}

Our sample consists of galaxies, identified in Paper 1, that have
characteristics intermediate between starbursts and AGNs based on
optical spectroscopy. We used the optical emission-line diagnostic
diagrams of Veilleux \& Osterbrock (1987), where the emission-line
ratios of [O~III]/H$\beta$, [N~II]/H$\alpha$, [S~II]/H$\alpha$ and
[O~I]/H$\alpha$ are compared, to find galaxies that are not clearly
identified as either a starburst or AGN. The emission-line ratios used
by Veilleux \& Osterbrock are generally good discriminators of
starbursts and AGNs, because each ratio consists of a forbidden line,
which is relatively strong in the partially ionised regions of AGNs,
compared with a hydrogen recombination line, which is excited by hot
stars formed in starbursts. The galaxies in our sample lie in an
intermediate region between the starburst and AGN groupings, either by
falling within $\pm$0.15 dex of the boundary line separating
starbursts and Seyferts, or by falling within the domain of starburst
galaxies in one diagnostic diagram and within the domain of AGNs in
another.  The sample was also constrained by a redshift upper limit of
$z = 0.035$ and declinations south of +24$^{\circ}$.  Full details of
the sample selection are given in Paper~1.

NIR spectroscopy, undertaken in Paper 1, showed that the ambiguous
nature of the galaxies was not limited to optical emission but that
NIR emission-line ratios also failed to distinguish the galaxies as
starburst or AGN. From this, and from further results in Paper 1, we
concluded that the galaxies are composites, containing both a
starburst and AGN component.

\section{Observations and Reductions}

\subsection{ATCA}

Radio continuum observations
were made in 1996 April 8--11, using the Australia Telescope Compact
Array in the 6A configuration, which utilises all six antennas and
covers east-west baselines from 337~m to 6~km. Observations were made
simultaneously at 6 and 3~cm, centred on frequencies of 4.800 and
8.640~GHz respectively, each spanning a bandwidth of 128~MHz. The
galaxies were observed in the time-efficient snapshot mode, whereby a
number of short observations, or cuts, of each galaxy are made over a
wide range of hour angles (Burgess \& Hunstead 1995). Between 10--20
cuts, each approximately 10 minutes, were made for every galaxy.
Unresolved phase calibrators were observed before and after every
galaxy observation; the primary flux density calibrator was
PKS~B1934$-$638.

Standard calibration and data reduction were performed using MIRIAD
(Sault {\it et~al.} 1995).  Natural weighting was applied to both
frequencies to maximise sensitivity.  The data were CLEANed and since
all fields contained at least one compact object, self-calibration was
used to provide additional corrections. Despite limited {\em uv}
coverage using the snapshot method, the noise level in each image was
close to theoretical, measuring 70--90~$\mu$Jy~beam$^{-1}$ rms. 
The resolution of the images ranged from 2$^{\prime\prime}$ to 
4$^{\prime\prime}$ FWHM at 3~cm and 4$^{\prime\prime}$ to 
8$^{\prime\prime}$ FWHM at 6~cm.

\subsection{PTI}

Radio interferometry observations were made at 13~cm using the
Parkes-Tidbinbilla Interferometer during two sessions in 1996, June
1--2 and October 19--20. Two bandpasses, centred on 2.290~GHz and
2.298~GHz, were observed simultaneously, each with bandwidths of
8~MHz.  The data were recorded in real-time using a microwave link
between the 64~m Parkes radio telescope and the 70~m telescope at
Tidbinbilla (part of the Canberra Deep Space Communication Complex),
providing a fringe spacing of 0.1$^{\prime\prime}$ at 2.3~GHz. Nineteen sample
galaxies were observed with PTI as well as three starburst galaxies
and three AGNs for comparison purposes. Unfortunately, it was not
possible to observe galaxies with RAs between
16$^{\rm{h}}$--20$^{\rm{h}}$ as the Tidbinbilla telescope was being
used to track the {\it Galileo\/} spacecraft.  Each galaxy was
observed at a single hour angle for approximately 20~minutes;
unresolved flat-spectrum calibrators were observed about every
2~hours.

The AIPS task FRING was used to search for detections in both delay
and fringe rate, by applying a Fourier transform fringe-search
technique.  In this way, we have been able to detect sources that are
offset from the phase and delay centre by plotting the correlated
intensity as a function of fringe frequency; fringe rate and delay
were then used to calculate source positions. Sources were detected up
to 3$^\prime$ away from the pointing centre. Furthermore, to ensure that
we did not miss any detections of galaxies within the sample due to
strong off-centre sources, FRING was run repeatedly, while decreasing
the area searched in delay and rate (see Section~4.3).

\section{Results}

\subsection{Radio Imaging}

Sixteen galaxies with declinations $< -15^{\circ}$ (see Table~1) were
observed with the ATCA, including four close galaxy pairs
(ESO~550-IG025, ESO~440-IG058, ESO~527-IG07 and ESO~343-IG013) and two
merging galaxies (MCG-02-33-098 and MCG-02-33-099). Optical spectra of
these galaxies, given in the literature, identify them as intermediate
between starbursts and AGNs (see Section~2 and Paper 1) with two
exceptions: the western nucleus of ESO~527-IG07 has no published
optical spectrum, and the merging galaxy MCG-02-33-099 has an optical
spectrum typical of a starburst.

In Figure~1 we present the ATCA 6~cm images overlaid as contours on
images from the Digitized Sky Survey (DSS), using the second epoch
(red-sensitive) DSS-II.\footnote{We note that Deutsch (1999) has found
positional errors $\sim 1^{\prime\prime}$ ($\approx$~1 pixel) in both RA and
Dec for some images in the prepublication release of the DSS-II.}
Of the 14 composite galaxies, 13 were detected with the ATCA and in
the majority of cases the emission is unresolved.  The only
non-detection was the northern nucleus of ESO~440-IG058, which has a
3$\sigma$ upper limit of 0.2~mJy.

Extended radio structure was found in two galaxies: the southern
galaxy in the pair ESO~440-IG58 and the southern galaxy in the pair
ESO~343-IG013. The latter pair, in particular, appears to represent a
current interaction with tidal tails being seen in optical images.  In
Figure~2 we present the ATCA 3~cm images overlaid as contours on the
DSS-II images for ESO~440-IG058 and ESO~343-IG013.  The increased
resolution of the 3~cm images shows that the extended radio emission
in ESO~440-IG058 south probably traces the star formation in the disk
of the galaxy.

Total flux densities for the galaxies, given in Table~2, were obtained
using the MIRIAD task SFIND, which performs a Gaussian profile fit.
We also include 20~cm flux densities from the NRAO VLA Sky Survey
(NVSS) (Condon {\it et~al.} 1998).
The NVSS images (45$^{\prime\prime}$ FWHM) are of lower resolution than the
ATCA images so, not surprisingly, the four galaxy pairs observed with
the ATCA (ESO~550-IG025, ESO~440-IG058, ESO~527-IG07 and
ESO~343-IG013) and the pair of merging galaxies (MCG-02-33-098 and
MCG-02-33-099) are detected as single sources by the NVSS.  For
ESO~440-IG058, ESO~527-IG07 and ESO~343-IG013 the 20~cm emission
appears to be centered on the dominant ATCA source.
For the non-interacting galaxies the 20~cm emission aligns closely
with the ATCA emission, except for ESO~374-IG032 where the NVSS image
includes an unrelated source detected at 6~cm with ATCA, lying
45$^{\prime\prime}$ south of the galaxy. For this galaxy, therefore, and for
the galaxy pairs, we have divided the 20~cm flux density into two
components in proportion to their 6~cm flux densities.

\subsection{Spectral Index}
\label{speci}

Ideally, to determine the spectral index, $\alpha$ (where $\rm{S}
\propto \nu^\alpha$), of the composite galaxies we require a scaled
array in which the baselines (in wavelengths) at 3 and 6~cm are
matched.  We approximate a scaled array for the 3~and 6~cm images by
applying natural weighting to both data sets, thereby giving more
significance to the shorter baselines, and by convolving the 3~cm
images to the corresponding 6~cm beam so that the images are brought
to the same resolution.

Although the spectral index is often determined between just two
frequencies, we draw on the NVSS data to expand the frequency
baseline.  In Table~2, we present the spectral index for each nucleus,
measured between 3--6~cm ($\alpha(6,3)$) and 6--20~cm
($\alpha(20,6)$).
Starburst and Seyfert galaxies typically have a spectral index of
$\alpha \approx -0.7$, although Seyferts are more likely to have
steeper spectra, with $\alpha < -1.1$.  (e.g., de Bruyn \& Wilson
1978; Sramek \& Weedman 1986; Edelson 1987).  The mean spectral index
for our galaxies, $\alpha(6,3) = -0.65 \pm 0.08$, is consistent with
the expected value for both starbursts and Seyferts.
The galaxy ESO~436-G026 has an unusually flat spectrum between 3 and
6~cm that steepens significantly towards 20~cm. We attribute this to
the presence of a flat-spectrum AGN core, as this galaxy was detected
with PTI and has a high core-to-total flux density ratio (see
Section~4.3).

\subsection{PTI}
\label{pti}

Unresolved radio emission (referred to here as a PTI `core') was
detected in five of the 20 composite galaxies observed with PTI. The
sensitivity limit of PTI, corresponding to a 3$\sigma$ detection, was
measured to be 0.9~mJy at 13~cm.  The flux densities of the five PTI
cores are given in Table~3, as well as the estimated total 13~cm flux
densities derived from the data in Table 2.  In addition to system
noise, the PTI core flux densities have nominal errors of $\approx$7\%
due to uncertainty in the antenna gains.  Studies of radio galaxies
involving multiple observations of individual sources (Slee {\it et~al.} 1994;
Morganti {\it et~al.} 1997) have shown that the correlated PTI flux density
can scatter by as much as 20\% due to time variability and/or source
structure. In our study only one galaxy, Mrk~1344, was observed twice,
in 1996 June and again in 1996 October, and the two measurements agree
to well within the combined calibration errors.

We now raise the question: Is our detection rate of 25\% consistent
with a sample of galaxies that contain AGNs? In the first instance we
compare this result with the detection rate found for the three
starburst galaxies and three Seyfert galaxies we observed for
comparison in this study. PTI cores were detected in two of the three
Seyfert galaxies but in none of the starbursts. This result is typical
of other studies: Norris {\it et~al.} (2000) detected only one of 74
starburst galaxies but obtained a PTI detection rate of 38\% from 221
Seyferts; Heisler {\it et~al.} (1998) detected none of three starbursts but
found PTI cores in 10 out of 11 Seyferts. We propose that the
detection rate for our composite sample, while not being exceptionally
high, is consistent with a sample of AGNs, assuming that PTI is
insensitive to the low brightness temperatures and diffuse structure
typical of starbursts.  This interpretation is apparently
contradicted by studies in which compact radio cores are
detected in many starburst galaxies.
One such study by Lonsdale {\it et~al.} (1993) found detection rates of
eight from 15 starbursts (53\%) and seven from 13 Seyferts (54\%) with
18~cm VLBI observations.  These high detection rates for starburst
galaxies, and also Seyferts, are perhaps not surprising as the
Lonsdale {\it et~al.} sample was preferentially selected from galaxies known
to contain compact radio cores with angular size $\le
0.25^{\prime\prime}$. The sample may also be subject to a luminosity selection
effect, as it was selected from among the most luminous members
of the IRAS Bright Galaxy Sample, with $\log(L_{FIR}/L_{\odot}) \geq
11.25$.

The detection of VLBI cores in many of the starburst galaxies in the
Lonsdale {\it et~al.} (1993) sample, combined with their strong IRAS
emission, may simply indicate a dominance of dust-obscured AGNs and
does not necessarily imply a starburst origin for the compact cores.
The alternative possibility, that a starburst-related process may
produce compact radio emission, comes from the discovery of a new
class of exceptionally luminous RSNe (Yin \& Heeschen 1991; Lonsdale
{\it et~al.} 1992; Wilkinson \& de Bruyn 1990).  
Smith {\it et~al.} (1998a) claim
that from 11 detections of starburst galaxies, 7 may be explained by
extremely luminous RSNe. However, the only galaxy to date where the
compact core can definitively be attributed to a starburst is the
ultraluminous IRAS galaxy (ULIRG), Arp 220, where high resolution
images have revealed numerous unresolved sources interpreted as RSNe
(Smith {\it et~al.} 1998b).

To investigate the possibility of RSNe in our sample we follow Smith
{\it et~al.} (1998b) and compare our PTI detections with the well studied
and unusually bright RSN, SN~1986J in the galaxy NGC~891 (Weiler,
Panagia \& Sramek 1990).  During 1986, SN~1986J peaked at 6~cm
reaching a flux density of $\sim$130~mJy. Using the (optically thick)
spectral index calculated at that time from 6~cm and 20~cm
observations of the supernova, we derive a corresponding 13~cm flux
density of $\sim$100~mJy.  Our sample galaxies are on average 10 times
more distant than NGC~891, so if a RSNe similar to 1986J existed in
our sample, it would be at a level of approximately 1~mJy and just
marginally detectable with PTI. Since 1986J is amongst the most
luminous RSNe ever discovered, the nature of the PTI detections
is not yet determined.

An interesting caveat in the debate over a starburst or AGN origin of
compact radio cores is given in the recent work of Kewley {\it et~al.}
(2000a).  In their study of warm infrared galaxies, they detected
compact radio cores in 8 from 10 (80\%) AGNs and 10 from 27 (37\%)
starbursts.  However, the core luminosities of the starburst
galaxies were lower than those in the AGNs, suggesting that the
starburst cores
may be radio-luminous supernovae. Comparing the luminosity of our five
PTI detections with those in the Kewley {\it et~al.} sample, we find that
three galaxies (ESO~550-IG025, MCG+00-29-23 and IRAS~12224-0624) are
consistent with AGN-like compact cores, while the remaining two
galaxies (ESO~436-G026 and Mrk 1344) are possibly
starburst-related. However, in the specific case of ESO~436-G026, the
flatness of its radio spectrum between 3--6~cm does favour the
presence of an AGN core.

We cannot tell from the radio data alone whether our PTI detections
are weak AGN cores or radio-luminous supernovae. However, we can
compare the flux density of the compact core (derived from the PTI
detections or 0.9~mJy limits for non-detections) with the total 13~cm
flux density (derived from the ATCA observations). These are used to
determine a core-to-total flux density ratio, given in Table~3. For
the composite galaxies we find on average that a compact core accounts
for $<10$\% of the total radio emission. Interestingly, the compact
radio cores in the reference Seyfert galaxies in Table 3 {\it also\/}
contribute typically $<10$\% of the total. We show in Section~4.7 that
a similar AGN contribution in the optical is sufficient to account for
the ambiguous optical emission line ratios of the composite galaxies.

\subsection{Supernova or AGN}

The PTI results indicate that most galaxies in our sample are not
AGN-dominated. Since it is generally assumed that the radio emission
from galaxies with starburst characteristics is predominantly
non-thermal emission from cosmic rays accelerated by supernova
remnants (SNRs), we follow Condon \& Yin (1990) and estimate the Type
II supernova rate from
\begin{equation}
\rm N_{SN} (yr^{-1}) = 7.7 \times 10^{-24} \, \nu^{\alpha} \, \, L_{NT},
\end{equation}
where $\rm L_{NT}$ is the non-thermal component of the radio
luminosity in WHz$^{-1}$, $\nu$ is the frequency in GHz and $\alpha$
is the non-thermal radio spectral index, taken here to be $\sim
-0.8$. The supernova rate can then be used to estimate the number of
ionising photons
\begin{equation}
\rm N_{UV} (s^{-1}) = 8.2 \times 10^{54} \, \, N_{SN}(yr^{-1})
\end{equation}
and, by assuming the average ionising flux from OB stars to be $ \rm
N_{UV} = 2 \times 10^{48}\, s^{-1}$ (Smith, Herter \& Haynes 1998), a
star-formation rate (SFR) can be derived and is given in Table~4.

Since the SFR is derived from the supernova rate, it is an indication
of the past SFR for massive stars in the galaxy, as the progenitors of
Type II SNe are massive stars ($\rm M > 8 \rm{M}_\odot$) with lifetimes 
$\sim 10^6$ years.  The SFR may be
somewhat over-estimated since we have assumed that all of the radio
emission is non-thermal. The empirical relationship between the
non-thermal and thermal radio emission (Condon \& Yin 1990) suggests
that 25\% of the total 4.85~GHz radio flux density may be thermal
(Smith, Herter \& Haynes 1998).

The galaxy NGC 1614 has an inferred supernova rate four times larger
than any other galaxy in the sample.  It has a high FIR luminosity,
and is usually classified as a starburst galaxy (Keto {\it et~al.} 1992).
However, based on our selection criteria, NGC~1614 has ambiguous
optical emission-line ratios (Veilleux {\it et~al.} 1995). Although Neff
{\it et~al.} (1990) find no direct evidence of an AGN in NGC~1614 they
suggest that the galaxy may be a precursor to a Seyfert (i.e., the AGN
has not yet turned on) due to the high luminosity they measure for the
nucleus of the galaxy.  No signs of an AGN have been found in other
studies of NGC~1614 (Lan\c{c}on {\it et~al.} 1996; Heisler {\it et~al.} 1999).

The median supernova rate from Table~4 is 0.2 yr$^{-1}$, which is high
compared with the rate of 0.03~yr$^{-1}$ found for Seyfert 2 galaxies
with circumnuclear star formation (Forbes \& Norris 1998). This
suggests a dominance of star formation within our sample. However, it
is surprising that Mrk~52, which was found to be dominated by star
formation in Paper 1, has an exceptionally weak SFR.

\subsection{FIR spectral energy distribution}
\label{firsed}

The heating mechanisms at work in galaxies can be investigated using
FIR colours. The majority of IRAS galaxies (70--80\%) appear to be
dominated by star formation, with the FIR emission being due to dust
heating from an intense starburst (Sanders \& Mirabel 1996; Leech
{\it et~al.} 1989).  This appears to hold even for the ULIRGs with $\rm
L_{FIR} > 10^{12} \; \rm L_\odot$ (Genzel {\it et~al.} 1998).  Mergers are
also common, with $\approx$~70\% of the ULIRGs involved in
interactions (Leech {\it et~al.} 1994). However, the proportion of mergers
does drop considerably for $\rm L_{FIR} < 10^{11} \; \rm L_\odot$
(Sanders \& Mirabel 1996). The extreme luminosities of some IRAS
galaxies, their extensive star formation, and the large fraction of
mergers have led to suggestions that these galaxies are presently in
a transient phase which may be important in the formation of an active
nucleus (Sanders {\it et~al.} 1988).

All the composite galaxies in our sample have published IRAS flux
densities and in Figure~3 we present three FIR colour-colour diagrams
which can be used to discriminate between starburst and Seyfert
excitation.  Included in these diagrams are a reddening line, extreme
mixing curve and empirical starburst line taken from Dopita {\it et~al.}
(1998). The reddening line was obtained by applying IR extinctions
given by Dwek {\it et~al.} (1997) to an average Seyfert 1, with flux density
normalised to 10 mJy at 100~$\mu$m. The empirical starburst line joins
the average flux values of a ``cool'' starburst to a ``hot'' starburst
taken from the data of Rush {\it et~al.} (1993), again with the flux density
normalised to 10 mJy at 100~$\mu$m. The extreme mixing line was
determined by adding a percentage of the ``cool'' starburst flux to
the non-reddened average Seyfert 1 value. Dopita {\it et~al.} (1998) make it
clear that any number of mixing curves can be added to these FIR
colour-colour diagrams but the mixing curve used here is an extreme
solution and, therefore, all objects are expected to lie within the
region bounded by the reddening line, the empirical starburst line and
the extreme mixing curve.

In Figure~3a the composite galaxies lie along the starburst line,
while in Figures~3b and 3c the composite galaxies are clustered around
the top end of the AGN/starburst mixing curve. This indicates that
star formation is dominating the FIR emission, with a starburst
contribution of $>$~90\%, consistent with the PTI results which place
the AGN contribution at $<$~10\%.

\subsection{FIR-radio correlation}

One of the tightest correlations for radio-quiet star-forming galaxies
is found between the FIR and radio luminosities. Although the
correlation has been known for two decades or more (Dickey \& Salpeter
1984; de~Jong {\it et~al.} 1985; Wunderlich, Klein \& Wielebinski 1987;
Condon 1992), its
origin is still not clear. Starburst activity is one likely
explanation: the massive OB stars formed within starbursts heat the
surrounding dust which re-radiates in the FIR, while the radio
emission comes about through the death of these massive stars as
supernovae that accelerate cosmic rays and in turn produce synchrotron
radiation (Helou \& Bicay 1993; Volk 1989).  The uncertainty in this
scenario is in the feedback mechanism required to link the FIR and
radio emission so tightly.

Starburst galaxies generally show a tighter FIR-radio correlation than
Seyferts (Roy {\it et~al.} 1998; Condon 1992; Norris {\it et~al.} 1988a), while
the more energetic AGNs, such as quasars and radio galaxies, do not fit
the correlation due to enhanced radio emission from the
AGN (Sopp \& Alexander 1991).  The FIR-radio correlation for the
present sample is shown in Figure~4, together with the fit from
de~Jong {\it et~al.} (1985). The composite galaxies are also seen to follow
the correlation. If the starburst origin for this correlation is
correct, then star formation is again seen to be the dominant process
in these galaxies.

\subsection{Comparison with optical and NIR spectroscopy}

\label{opt}

As mentioned in Section~2, the sample of composite galaxies was chosen
on the basis of the Veilleux \& Osterbrock (1987) optical emission-line 
diagnostic diagrams. These diagrams are presented in Figure~5 and 
include starburst and AGN data from the literature, and also two
photoionisation models developed in Paper~1 using MAPPINGS~II (see
Sutherland \& Dopita 1993). The starburst model is based on the
stellar atmosphere models of Hummer \& Mihalas (1970) with a stellar
temperature of 40~000~K, while the AGN model was created from a
power-law ionising spectrum of the form $f_\nu \sim \nu^{\alpha}$,
with $\alpha=-1.5$. Other model parameters include a hydrogen density
of 10$^3$~cm$^{-3}$ and solar metallicity.

The galaxies that were detected with PTI do not appear to occupy any
special position within Figure~5. However, we have also included on
the diagnostic diagrams a mixing line from the low-ionisation
starburst galaxies to the high-ionisation AGNs. The end points for the
mixing line were determined by taking a median of the emission-line
ratios for starbursts and AGNs (with a cut-off of $\rm
\log([OIII]/H\alpha) > 0.5$ applied to the AGN ratios to omit LINERS)
taken from Veilleux \& Osterbrock (1987 and references therein) and
Veilleux {\it et~al.} (1995).  It becomes evident that an AGN contribution
of $\approx$~10\% of the total emission of the galaxy can still
account for the borderline position of the composite galaxies within
the diagnostic diagrams.  In fact, it has been recently found that the
optical diagnostic diagrams are especially sensitive to the presence
of an AGN.  Kewley {\it et~al.} (2000b) applied a mixing line from their
photoionisation (or starburst) model to their shock (or AGN) model and
found that objects with as little as 20\% AGN contribution to their
energy budget may be classified as AGN in the optical diagnostic
diagrams.

Our study of NIR diagnostic diagrams in Paper 1 found that the line
ratio diagram of [Fe~II]/Pa$\beta$ vs [O~I]/H$\alpha$ was the most
useful for distinguishing starbursts and Seyferts.  This diagram is
shown in Figure~6 with the same starburst and power-law models
presented in Figure~5 and with the composite galaxies again
differentiated by the PTI detections. The two galaxies detected with
PTI, and for which we have NIR data, have the largest
[Fe~II](1.25$\mu$m)/Pa$\beta$ ratios and therefore are most likely to
be classified as AGNs from this diagram (see Paper 1).

\section{Conclusions}

We have presented arc-second radio images and long-baseline radio
interferometer observations to extend our study of  
composite galaxies with ambiguous optical emission-line
ratios. Extended radio morphology in radio-quiet galaxies is usually
attributed to starbursts and we found that only two galaxies,
ESO~440-IG058 south and ESO~343-IG013 south, showed extended emission
with the ATCA. The radio continuum images were also used to derive
spectral indices for the galaxies, with the average being $\alpha(6,3)
= -0.65 \pm 0.08$ which is typical of both starbursts and Seyferts.

Our interferometry observations revealed compact cores in five of the
20 galaxies observed, which may be due either to a weak ($<10$\% of
total flux density) AGN component or a radio-luminous supernova.  The
majority of non-detections for the sample does not rule out the
existence of an AGN core but points to the starburst being the
dominant flux density contribution within the galaxies.  A similar
result was found upon examining the FIR emission which we also found
to be dominated by star formation.

From our radio observations and examination of the FIR fluxes we
conclude that star formation dominates the composite galaxies, and we
attribute only a small fraction of the emission ($\le$~10\%) to an AGN
component. However, we have shown that an AGN contribution as small as
this is still sufficient to identify the galaxies as composites within
the optical diagnostic diagrams.

\acknowledgments

We thank Edward King and staff at the Tidbinbilla Tracking Station for
help with the PTI observations. We also thank the referee for
insightful comments which greatly improved this paper. TLH
acknowledges receipt of an Australian Postgraduate Award, and thanks
Museum Victoria and Rachel Webster for support during the writing of 
this paper.  RWH
acknowledges funding from the Australian Research Council.  The
Australia Telescope Compact Array and the Parkes Telescope are part of
the Australia Telescope which is funded by the Commonwealth of
Australia for operation as a National Facility managed by CSIRO.  This
research has made use of the NASA Astrophysics Data System service and
of the NASA/IPAC Extragalactic Database which is operated by the Jet
Propulsion Laboratory, California Institute of Technology, under
contract with the National Aeronautics and Space Administration. This
research has also made use of the Digitized Sky Surveys (DSS) produced
at the Space Telescope Science Institute under U.S. Government grant
NAG W-2166.

\clearpage
\newpage


\begin{deluxetable} {lccccc}
\tablecaption{Observing Log \label{obs}}
\tabletypesize{\normalsize}
\tablewidth{0pt}
\tablehead{
\colhead{Galaxy} &
\multicolumn{2}{c}{Optical position} &
\colhead{z} &
\colhead{ATCA} & 
\colhead{PTI} \nl
&\colhead{RA(J2000)} &
\colhead{Dec(J2000)} &
& \colhead{Int. Time} & \colhead{Obs} \nl
&&&& \colhead{(min)} &  
}
\tablecolumns{6}
\startdata
NGC 232 & 00 42 45.9 & $-$23 33 36 & 0.023 & 228 & Y  \nl
NGC 1204 & 03 04 40.4 & $-$12 20 28 & 0.015 & \nodata & Y \nl
ESO 550-IG025 N & 04 21 19.9 & $-$18 48 39 & 0.032 & 139 & Y \nl
ESO 550-IG025 S & 04 21 20.0 & $-$18 48 56 & 0.032 & 139 & Y \nl
NGC 1614 & 04 33 59.8 & $-$08 34 44 & 0.016 & \nodata & Y \nl
ESO 374-IG032 & 10 06 04.5 & $-$33 53 08 & 0.034 & 108 & Y  \nl 
IRAS 10057$-$3343 & 10 07 59.1 & $-$33 58 07 & 0.034 & \phn98 & Y \nl 
ESO 500-G034 & 10 24 31.4 & $-$23 33 11 & 0.013 & 101 & Y \nl
ESO 436-G026 & 10 28 42.7 & $-$31 02 18 & 0.014 & 130 & Y \nl
MCG+00-29-23 & 11 21 11.7 & $-$02 59 03 & 0.025 & \nodata & N \nl
Mrk 739 (NGC 3758) & 11 36 29.1 & +21 35 48 & 0.030 & \nodata & Y  \nl
ESO 440-IG058 N & 12 06 51.7 & $-$31 56 47 & 0.023 & 167 & Y  \nl 
ESO 440-IG058 S & 12 06 51.8 & $-$31 56 58 & 0.023 & 167 & Y  \nl 
IRAS 12224$-$0624 & 12 25 04.0 & $-$06 40 53 & 0.026 & \nodata & Y  \nl
Mrk 52 (NGC 4385) & 12 25 42.6 & +00 34 23 & 0.0071 & \nodata & Y  \nl
MCG-02-33-098\tablenotemark{a} & 13 02 19.9 & $-$15 46 06 & 0.017 & 175 & Y \nl
Mrk 1344 (NGC 4990) & 13 09 17.2 & $-$05 16 23 & 0.011 & \nodata & Y \nl
NGC 5713 N\tablenotemark{b} & 14 40 11.1 & $-$00 17 02 & 0.0073 & \nodata & Y \nl
NGC 5719 & 14 40 56.7 & $-$00 18 58 & 0.0058 & \nodata & Y \nl
NGC 5937 & 15 30 46.6 & $-$02 49 32 & 0.0094 & \nodata & Y \nl
ESO 527-IG07 E\tablenotemark{c} & 20 04 31.3 & $-$26 25 40  & 0.035 & 175 & N  \nl
ESO 343-IG013 N & 21 36 11.1 & $-$38 32 33 & 0.019 & 197 & Y  \nl
ESO 343-IG013 S & 21 36 11.0 & $-$38 32 42 & 0.019 & 197 & Y  \nl
ESO 602-G025 & 22 31 25.3 & $-$19 02 05 & 0.025 & 208 & Y  \nl
\cutinhead{Starbursts}
AKN 232 & 10 07 38.8 & +17 06 02 & 0.026 & \nodata & Y  \nl
Mrk 717 (IC 2551) & 10 10 40.3 & +24 24 51 & 0.021 & \nodata & Y \nl
Mrk 529 (NGC 7532) & 23 14 22.2 & $-$02 43 39 &  0.010 & \nodata & Y \nl
\cutinhead{Seyferts}
Mrk 955  & 00 37 35.8 & +00 16 51& 0.035 & \nodata & Y \nl
NGC 3627 (M66) & 11 20 15.0 & +12 59 30 & 0.0024 & \nodata & Y \nl
NGC 4569 (M90) & 12 36 49.8 & +13 09 46 & 0.00078 & \nodata & Y  \nl
\enddata
\tablenotetext{a}{MCG-02-33-098 is the western galaxy of an interacting
pair. The eastern galaxy, MCG-02-33-099 has an optical
spectrum typical of a starburst.}
\tablenotetext{b}{The southern nucleus of NGC 5713 has an optical spectrum
typical of a starburst.}
\tablenotetext{c}{The western nucleus of ESO 527-IG07 has no published
optical spectrum.}
\end{deluxetable}

\begin{deluxetable} {lccccccc}
\tablecaption{Radio continuum flux densities \& spectral indices \label{flux}}
\tabletypesize{\small}
\tablewidth{0pt}
\tablehead{
\colhead{Galaxy} &
\multicolumn{2}{c}{Radio position} &
\colhead{3~cm\tablenotemark{a}} &
\colhead{6~cm} &
\colhead{20~cm\tablenotemark{b}} &
\colhead{$\alpha$(6,3)} &
\colhead{$\alpha$(20,6)} \nl
&
\colhead{RA(J2000)} &
\colhead{Dec(J2000)} &
\colhead{(mJy)} & \colhead{(mJy)} & \colhead{(mJy)}  & & \nl
}
\tablecolumns{8}
\startdata
NGC 232 & 00 42 45.8 & $-$23 33 41 & 14.3 & 22.4 & 60.6 &$-$0.76 & $-$0.81 \nl
NGC 1204 & 03 04 40.0 & $-$12 20 30 & \nodata & \nodata & 26.4 & \nodata & \nodata \nl
ESO 550-IG025 N & 04 21 20.0 & $-$18 48 39 & 7.2 & 10.5 & 28.3\tablenotemark{c} & $-$0.64 & $-$0.80 \nl
ESO 550-IG025 S & 04 21 20.0 & $-$18 48 57 & 3.2 & 5.2 & 13.3\tablenotemark{c} & $-$0.83 & $-$0.76 \nl
NGC 1614 & 04 34 00.0 & $-$08 34 45 & \nodata & 63\tablenotemark{d} & 138.2 &
\nodata & $-$0.64 \nl
ESO 374-IG032 & 10 06 04.6 & $-$33 53 06 & 3.7 & 5.5 & 18.7\tablenotemark{c} &
$-$0.67 & $-$0.99 \nl
IRAS 10057$-$3343 & 10 07 59.0 & $-$33 58 06 & 3.7 & 5.3 & 13.2 & $-$0.61 & $-$0.74 \nl
ESO 500-G034 & 10 24 31.5 & $-$23 33 11 & 15.7 & 22.0 & 57.8  & $-$0.57 & $-$0.78 \nl
ESO 436-G026 & 10 28 42.9 & $-$31 02 17 & 2.9 & 3.0 & 11.9  & $-$0.01 & $-$1.12 \nl
Mrk 739 & 11 36 29.2 & +21 35 49 & \nodata & \nodata & 11.2 & \nodata & \nodata  \nl
ESO 440-IG058 N & \nodata & \nodata & \llap{$<$}0.3\tablenotemark{e} & 
\llap{$<$}0.2\tablenotemark{e} & \nodata & \nodata & \nodata \nl
ESO 440-IG058 S & 12 06 51.9 & $-$31 56 59 & 9.7 & 18.4 & 52.1 & $-$1.09 & $-$0.84  \nl
IRAS 12224$-$0624 & 12 25 04.1 & $-$06 40 54 & \nodata & \nodata & 11.2 & \nodata & \nodata   \nl
Mrk 52 & 12 25 42.8 & +00 34 23 & \nodata & \nodata & 14.8 & \nodata & \nodata  \nl
MCG-02-33-098 & 13 02 19.7 & $-$15 46 04 & 5.6 & 7.7 & 22.1\tablenotemark{c} & $-$0.54 & $-$0.86 \nl
Mrk 1344 & 13 09 17.3 & $-$05 16 21 & \nodata & \nodata & 29.9  & \nodata & \nodata  \nl
NGC 5713 & 14 40 11.3 & $-$00 17 26 & \nodata & 81\tablenotemark{d} & 159.9  & \nodata & $-$0.55 \nl
NGC 5719 & 14 40 56.3 & $-$00 19 07 & \nodata & \nodata & 58.7 & \nodata & \nodata  \nl
NGC 5937 & 15 30 46.1 & $-$02 49 47 & \nodata & \nodata & 114.0& \nodata & \nodata   \nl
ESO 527-IG07 E & 20 04 31.0 & $-$26 25 39 & 4.0 & 6.1 & 12.8\tablenotemark{c} & $-$0.72 & $-$0.60  \nl
ESO 343-IG013 N & 21 36 10.9 & $-$38 32 33 & 2.6 & 3.2 & 7.9\tablenotemark{c}& $-$0.35 & $-$0.73  \nl
ESO 343-IG013 S & 21 36 10.6 & $-$38 32 43 & 5.6 & 7.8 & 18.0\tablenotemark{c}& $-$0.56 & $-$0.68  \nl
ESO 602-G025 & 22 31 25.5 & $-$19 02 04 & 7.4 & 14.2 & 46.5 & $-$1.11 & $-$0.96 \nl
\cutinhead{Starbursts}
AKN 232 & 10 07 38.8 & +17 06 02 & \nodata & \nodata & \nodata & \nodata & \nodata \nl
Mrk 717 (IC 2551) & 10 10 40.3 & +24 24 50 & \nodata & \nodata & 26.0 & \nodata & \nodata  \nl
Mrk 529 (NGC 7532) & 23 14 22.3 & $-$02 43 42 & \nodata  & \nodata & 11.6 & \nodata & \nodata  \nl
\cutinhead{Seyferts}
Mrk 955  & 00 37 36.0 & +00 16 53& \nodata & \nodata & 8.4 & \nodata & \nodata  \nl
NGC 3627 (M66) & 11 20 16.8 & +12 58 46 &\nodata  & 141\tablenotemark{f} & 466\tablenotemark{g} & \nodata & \nodata  \nl
NGC 4569 (M90) & 12 36 49.6 & +13 09 57 & \nodata & \nodata & 73.1 & \nodata & \nodata  \nl
\enddata
\tablenotetext{a}{The 3~cm image has been convolved to the 6~cm beamwidth to 
approximate a scaled array.}
\tablenotetext{b}{The 20~cm flux densities are taken from the NVSS (Condon {\em
et~al.} 1998)}
\tablenotetext{c}{The NVSS source is blended, and the 6~cm ATCA
flux densities have been used to determine the relative contributions at 20~cm.}
\tablenotetext{d}{Griffith {\em et~al.} (1995)}
\tablenotetext{e}{3$\sigma$ limit}
\tablenotetext{f}{Gregory {\em et~al.} (1995)}
\tablenotetext{g}{This is a complicated region where three NVSS sources are blended.}
\end{deluxetable}

\begin{deluxetable} {lccccr}
\tablecaption{PTI detections and upper limits \label{ptired}}
\tabletypesize{\normalsize}
\tablewidth{0pt}
\tablehead{
\colhead{Galaxy} & \multicolumn{2}{c}{Position of PTI detection} &
\colhead{PTI flux} &
\colhead{Total 13~cm} & 
\colhead{core/} \nl
& \colhead{RA(J2000)} & \colhead{Dec(J2000)} & \colhead{density} & 
\colhead{flux density\tablenotemark{a}} & 
\colhead{total}  \nl
\colhead{} & & & \colhead{(mJy)} & \colhead{(mJy)} & \nl
}
\tablecolumns{6}
\startdata
NGC 232 & \nodata & \nodata & \llap{$<$}0.9 & 39.2 & \llap{$<$}2\%  \nl
NGC 1204 & \nodata & \nodata &  \llap{$<$}0.9& 19.6 & \llap{$<$}5\%  \nl
ESO 550-IG025 & 04 21 18.3 & $-$18 48 49 & 1.9 & 26.2 & 7\%  \nl
NGC 1614 & \nodata & \nodata &  \llap{$<$}0.9& 102 & \llap{$<$}0.9\%  \nl
ESO 374-IG032 & \nodata & \nodata &  \llap{$<$}0.9& 9.1 & \llap{$<$}10\% \nl
IRAS 10057$-$3343 & \nodata & \nodata &  \llap{$<$}0.9& 8.1 & \llap{$<$}11\%  \nl
ESO 500-G034 & \nodata & \nodata &  \llap{$<$}0.9& 33.5 & \llap{$<$}3\% \nl
ESO 436-G026 & 10 28 43.3 & $-$31 02 22 & 1.0 & 3.0 & 33\%  \nl
MCG+00-29-23 & \nodata & \nodata & 5\tablenotemark{b} & \nodata & \nodata \nl
Mrk 739 & \nodata & \nodata &  \llap{$<$}0.9& 8.3 & \llap{$<$}11\% \nl
ESO 440-IG058 & \nodata & \nodata &  \llap{$<$}0.9& 41.3 & \llap{$<$}2\%  \nl
IRAS 12224$-$0624 & 12 25 05.9 & $-$06 40 29 & 3.1 & 8.3 & 37\% \nl
Mrk 52 & \nodata & \nodata &  \llap{$<$}0.9& 11.0 & \llap{$<$}8\% \nl
MCG-02-33-098 & \nodata & \nodata &  \llap{$<$}0.9& 11.5 & \llap{$<$}8\% \nl
Mrk 1344 & 13 09 17.3 & $-$05 16 21 & 3.5 & 22.2 & 16\% \nl
NGC 5713 N & \nodata & \nodata &  \llap{$<$}0.9& 119 & \llap{$<$}0.8\%  \nl
NGC 5719 & \nodata & \nodata &  \llap{$<$}0.9& 43.5 & \llap{$<$}2\%  \nl
NGC 5937 & \nodata & \nodata &  \llap{$<$}0.9& 84.5 & \llap{$<$}1\% \nl
ESO 343-IG013 & \nodata & \nodata &  \llap{$<$}0.9& 15.9 & \llap{$<$}6\%  \nl
ESO 602-G025 & \nodata & \nodata &  \llap{$<$}0.9& 31.9 & \llap{$<$}3\% \nl
\cutinhead{Starbursts}
AKN 232 & \nodata & \nodata & \llap{$<$}0.9 & \nodata & \nodata \nl
Mrk 717 & \nodata & \nodata & \llap{$<$}0.9 & 19.2 & \llap{$<$}5\% \nl
Mrk 529 & \nodata & \nodata &  \llap{$<$}0.9& 8.6 & \llap{$<$}10\% \nl
\cutinhead{Seyferts}
Mrk 955  & \nodata & \nodata &  \llap{$<$}0.9& 6.2 & \llap{$<$}15\%  \nl
NGC 3627 & 11 20 15.5 & +12 59 20 & 2.0 & 345 & 0.6\%  \nl
NGC 4569 & 12 36 46.6 & +13 09 12 & 1.1 & 54.2 & 2\% \nl
\enddata
\tablenotetext{a}{Derived from 6~cm flux density and $\alpha$(6,3) spectral
index; however, if no ATCA data was available we used the
20~cm fluxes and assumed a spectral index of $-$0.7.}
\tablenotetext{b}{Norris {\it et~al.} (1990)}
\end{deluxetable}

\begin{deluxetable} {lccc}
\tablecaption{Derived Supernova and Star Formation Rates \label{sfr}}
\tabletypesize{\normalsize}
\tablewidth{0pt}
\tablehead{
\colhead{Galaxy} &
\colhead{$\rm N_{SN}$} & 
\colhead{$\rm SFR(M > 8 M_\odot)$}  \nl
\colhead{} &
\colhead{$\rm yr^{-1}$} &
\colhead{M$_\odot$ yr$^{-1}$}  &
\nl
}
\tablecolumns{3}
\startdata
NGC 232 & 0.62 & 3.1 \nl
NGC 1204 & 0.12 & 0.6  \nl
ESO 550-IG025 & 0.88 & 4.4  \nl 
NGC 1614 & 3.39 & 16.9  \nl
ESO 374-IG032 & 0.33 & 1.7 \nl
IRAS 10057$-$3343 & 0.32 & 1.6 \nl
ESO 500-G034 & 0.19 & 1.0  \nl
ESO 436-G026 & 0.03 & 0.2  \nl
MCG+00-29-023 & \nodata & \nodata  \nl
Mrk 739 & 0.20 & 1.0 \nl
ESO 440-IG058 & 0.51 & 2.5  \nl
IRAS 12224$-$0624 & 0.15 & 0.8  \nl
Mrk 52 & 0.01 & 0.07   \nl
MCG-02-33-098 & 0.19 & 0.9 \nl
Mrk 1344 & 0.07 & 0.4  \nl
NGC 5713 & 0.21 & 1.0 \nl
NGC 5719 & 0.59 & 0.3  \nl
NGC 5937 & 0.19 & 0.9 \nl
ESO 527-IG07 E & 0.61 & 3.1 \nl
ESO 343-IG013 & 0.22 & 1.1  \nl
ESO 602-G025 & 0.46 & 2.3  \nl
\enddata
\end{deluxetable}

\onecolumn

\begin{figure} 
\figurenum{1}
\epsscale{0.5}
\begin{tabular} {ccc}
\plotone{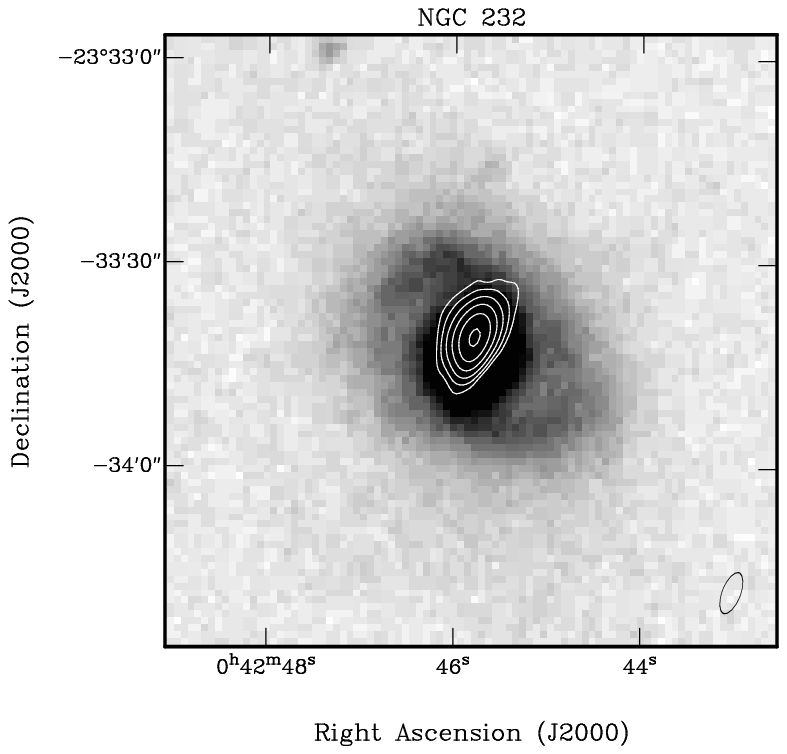} & \plotone{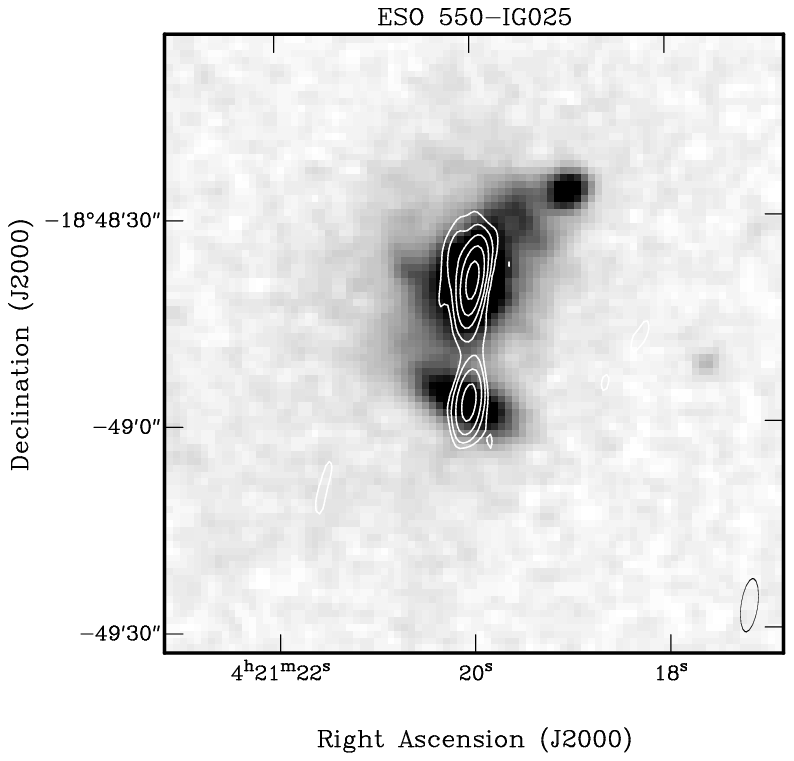} & \\
\plotone{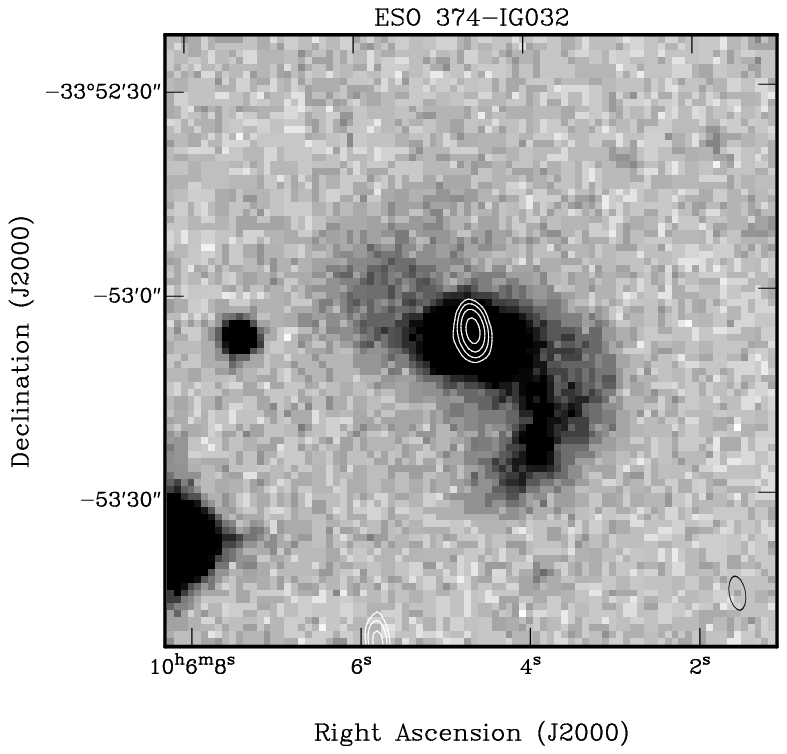} & \plotone{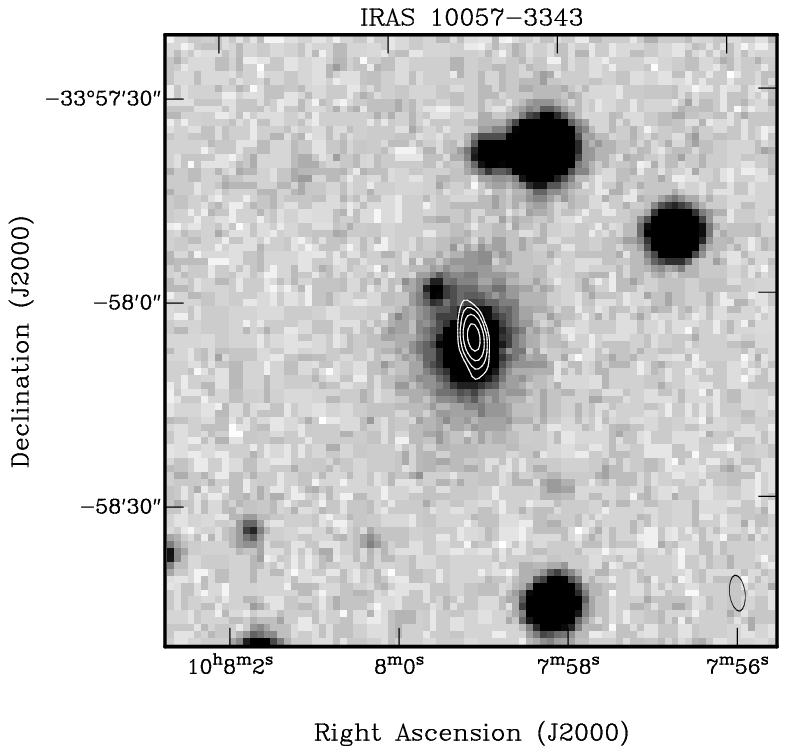} &  \\
\end{tabular}
\caption{DSS images (greyscale) overlaid with 6~cm
ATCA images (contours) for all the composite galaxies. The contours
are at 0.4, 0.8, 1.6, 3.2, 6.4 and 12.8 mJy~beam$^{-1}$. The
synthesised half-power beam-shape is shown as an ellipse in the lower
right corner of each image. The red-sensitive DSS-II has been used
for all overlays except MCG-02-33-098 which uses the blue DSS-I.}
\end{figure}
\clearpage

\begin{figure} 
\figurenum{1}
\epsscale{0.5}
\begin{tabular} {ccc}
\plotone{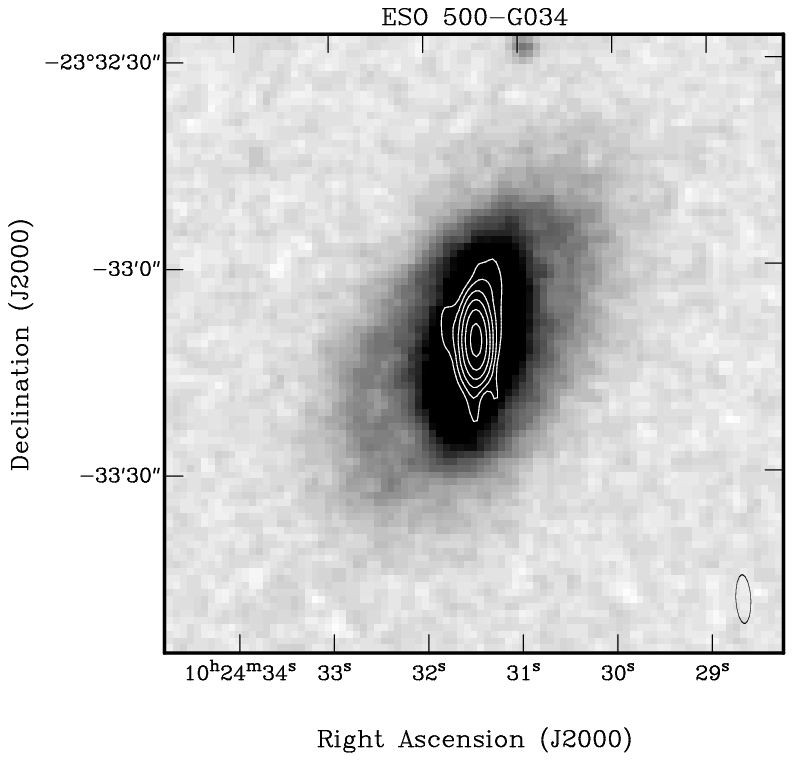} & \plotone{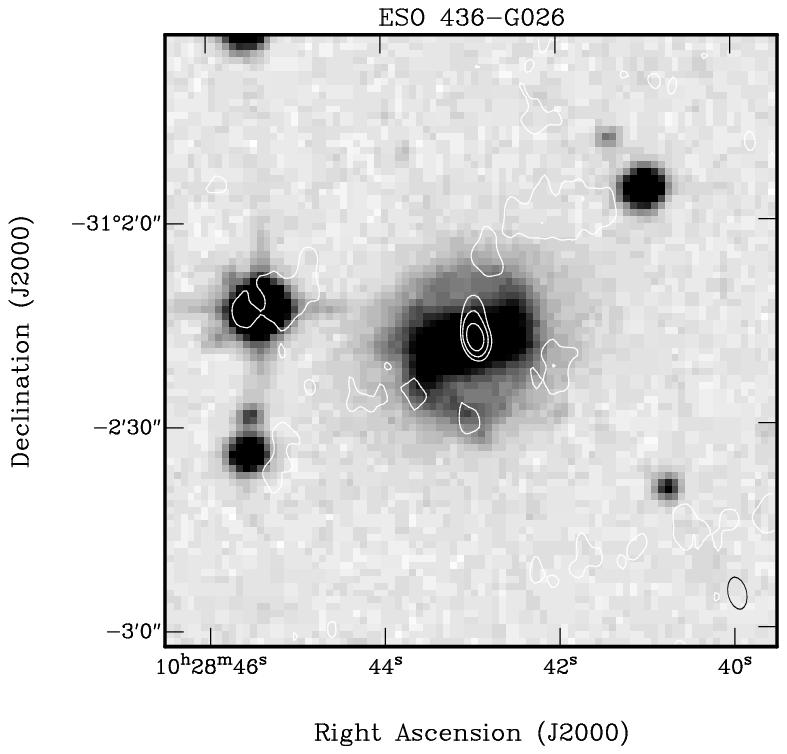} & \\
\plotone{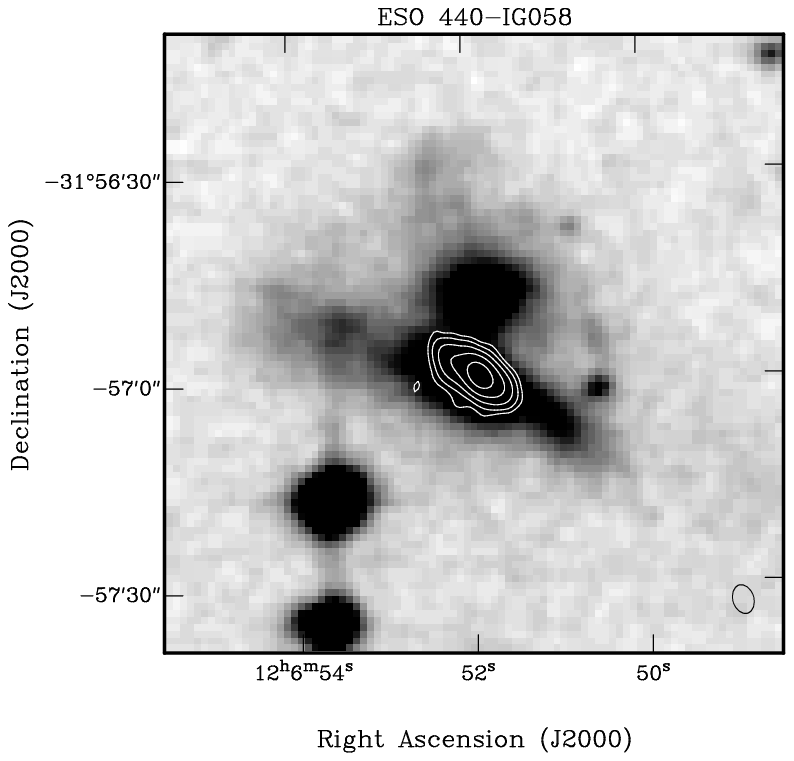} & \plotone{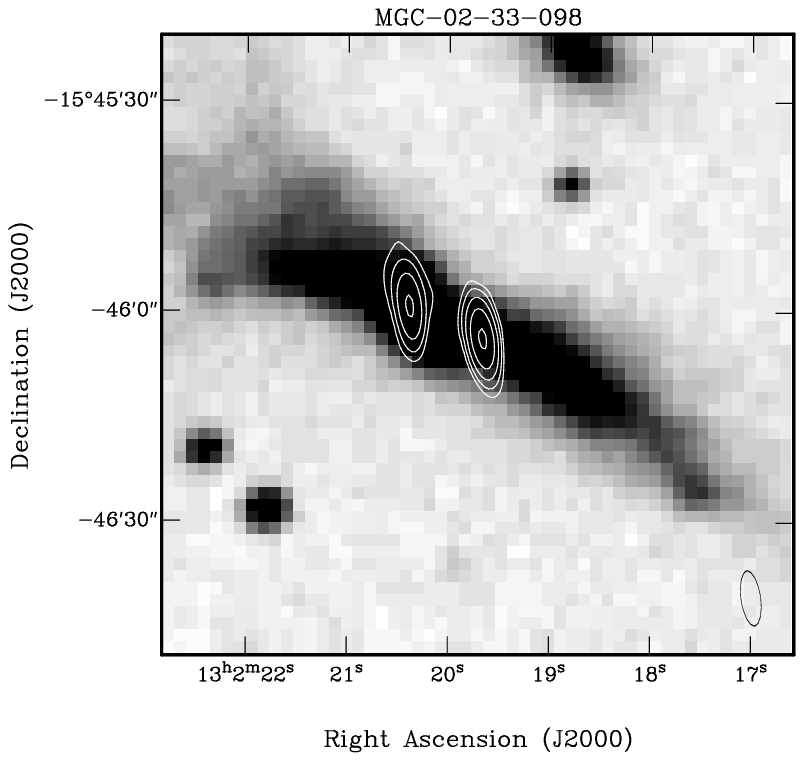} & \\
\end{tabular}
\caption{\em cont.}
\end{figure}
\clearpage

\begin{figure} 
\figurenum{1}
\epsscale{0.5}
\begin{tabular} {ccc}
\plotone{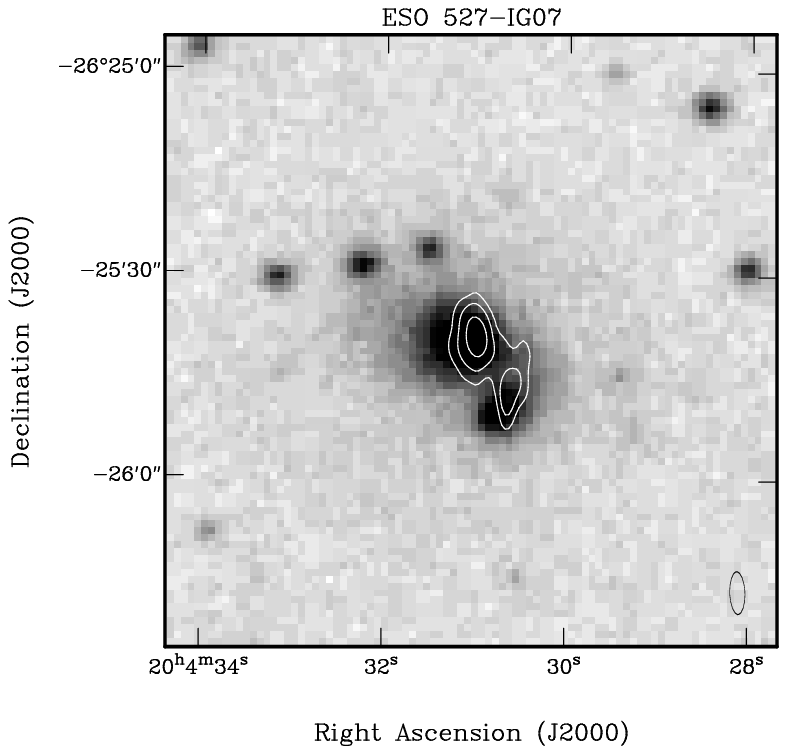} & \plotone{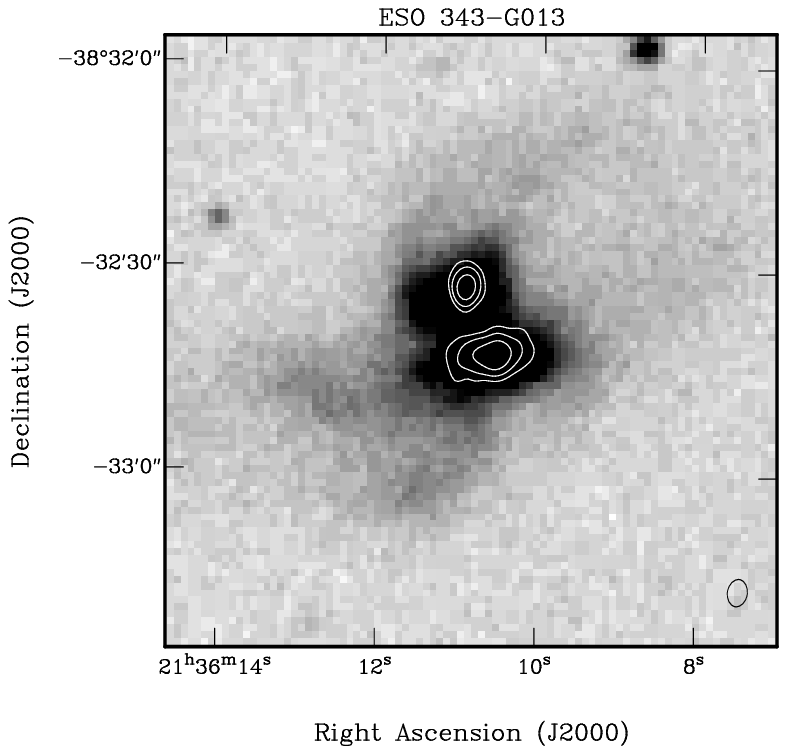} & \\
\plotone{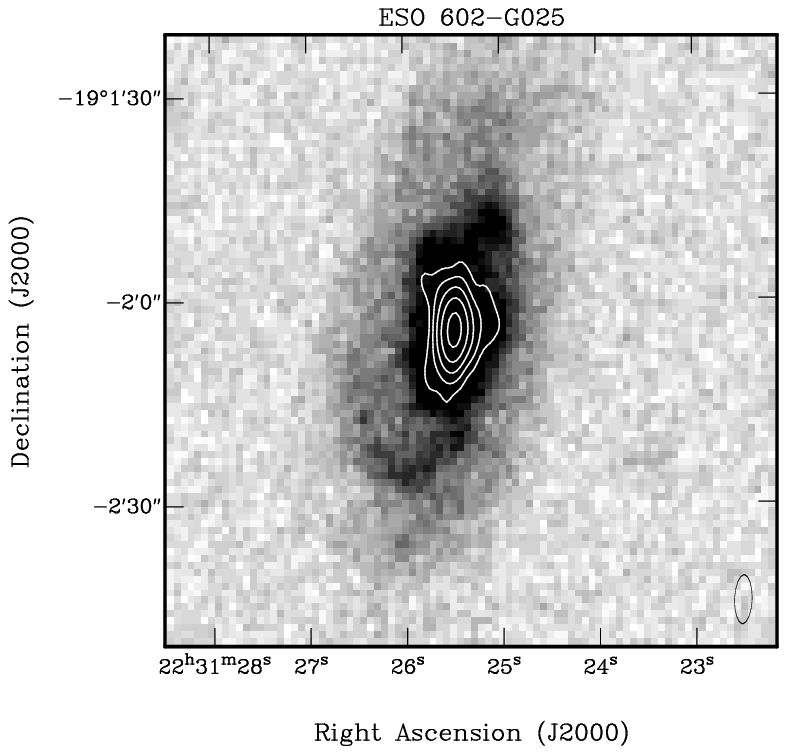} & & \\
\end{tabular}
\caption{\em cont.} 
\end{figure}
\clearpage

\begin{figure} 
\figurenum{2}
\epsscale{0.5}
\begin{tabular} {ccc}
\plotone{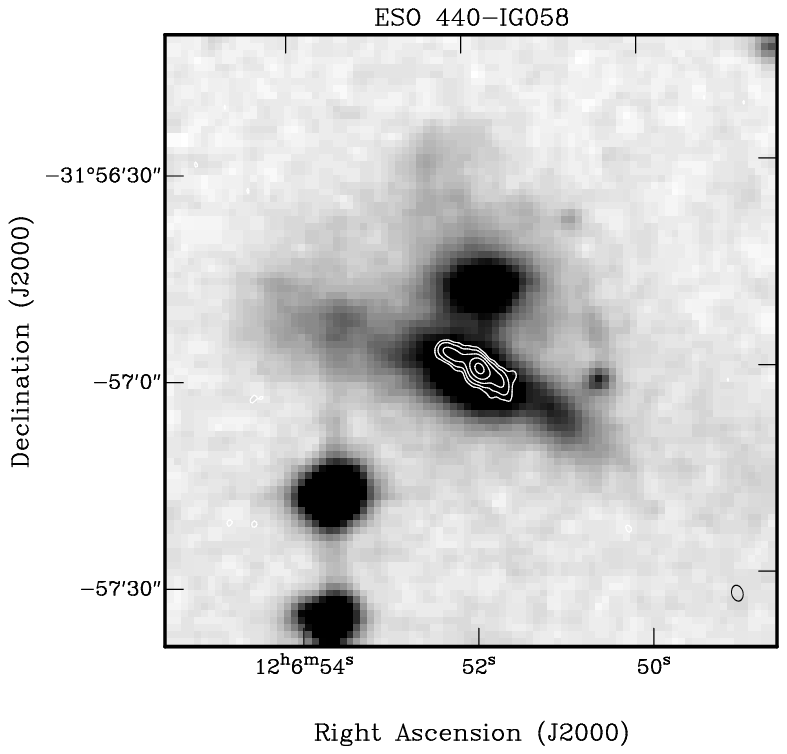} & \plotone{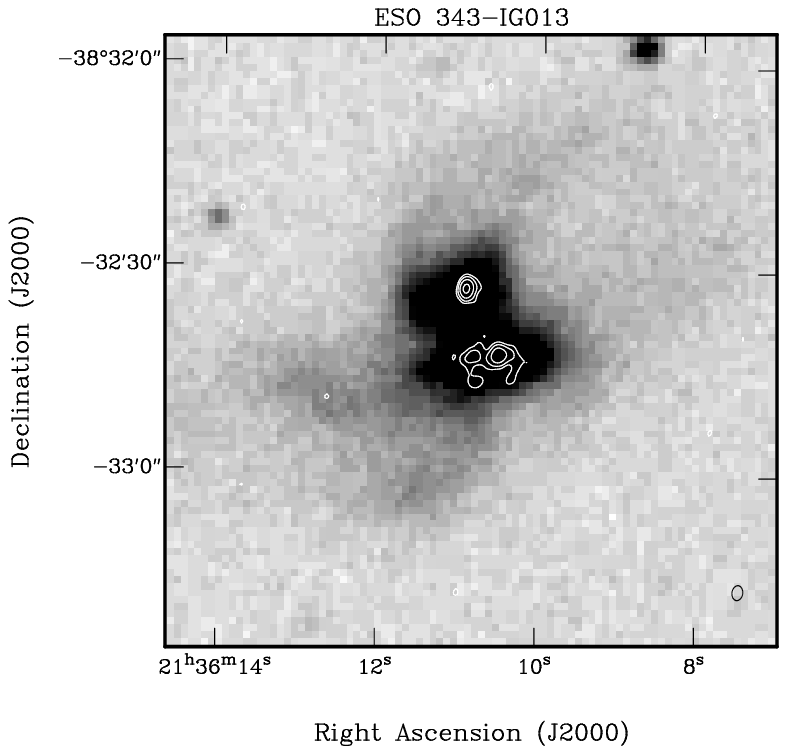} & \\
\end{tabular}
\caption[hill.fig2.ps] {DSS-II images (greyscale) overlaid with
3~cm ATCA images (contours) for the two galaxies in the sample with
extended emission. The contours are at 0.2, 0.4, 0.8, 1.6, 3.2 mJy
beam$^{-1}$. The synthesised half-power beam-shape is shown as an
ellipse in the lower right corner of each image.}
\end{figure}
\clearpage

\begin{figure}
\epsscale{0.50}
\figurenum{3}
\begin{tabular} {ccc}
\plotone{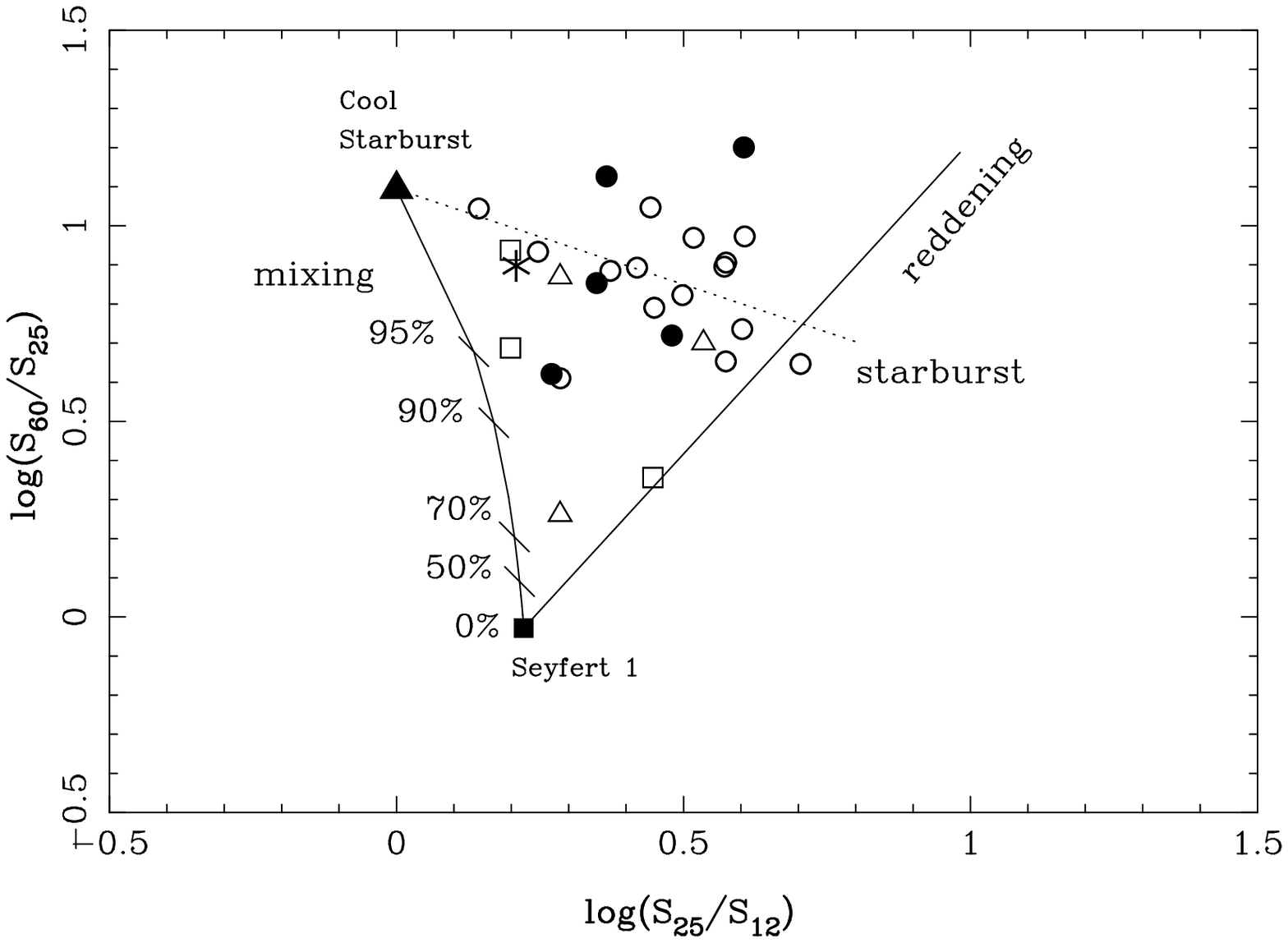} & \plotone{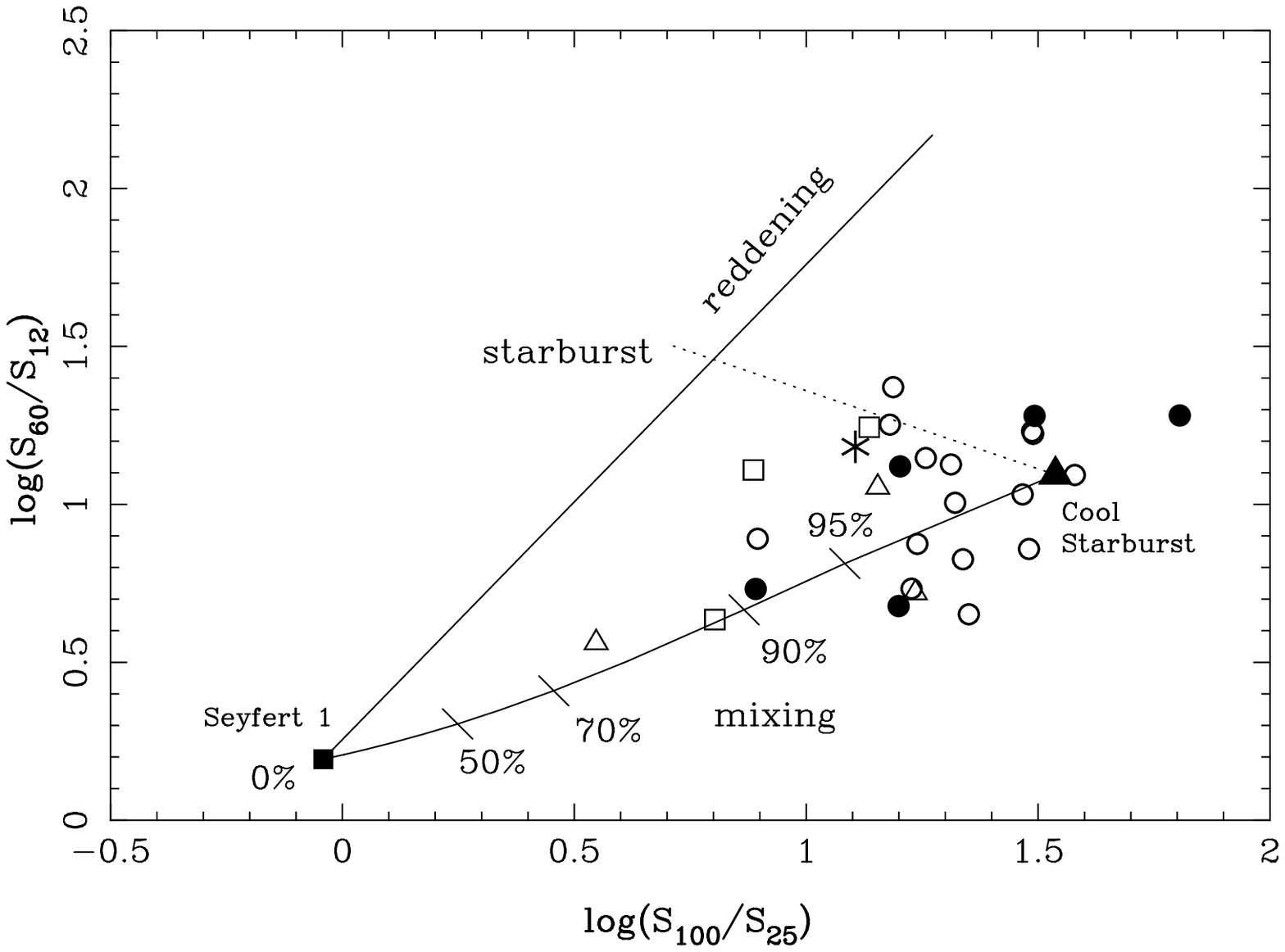} & \\
\plotone{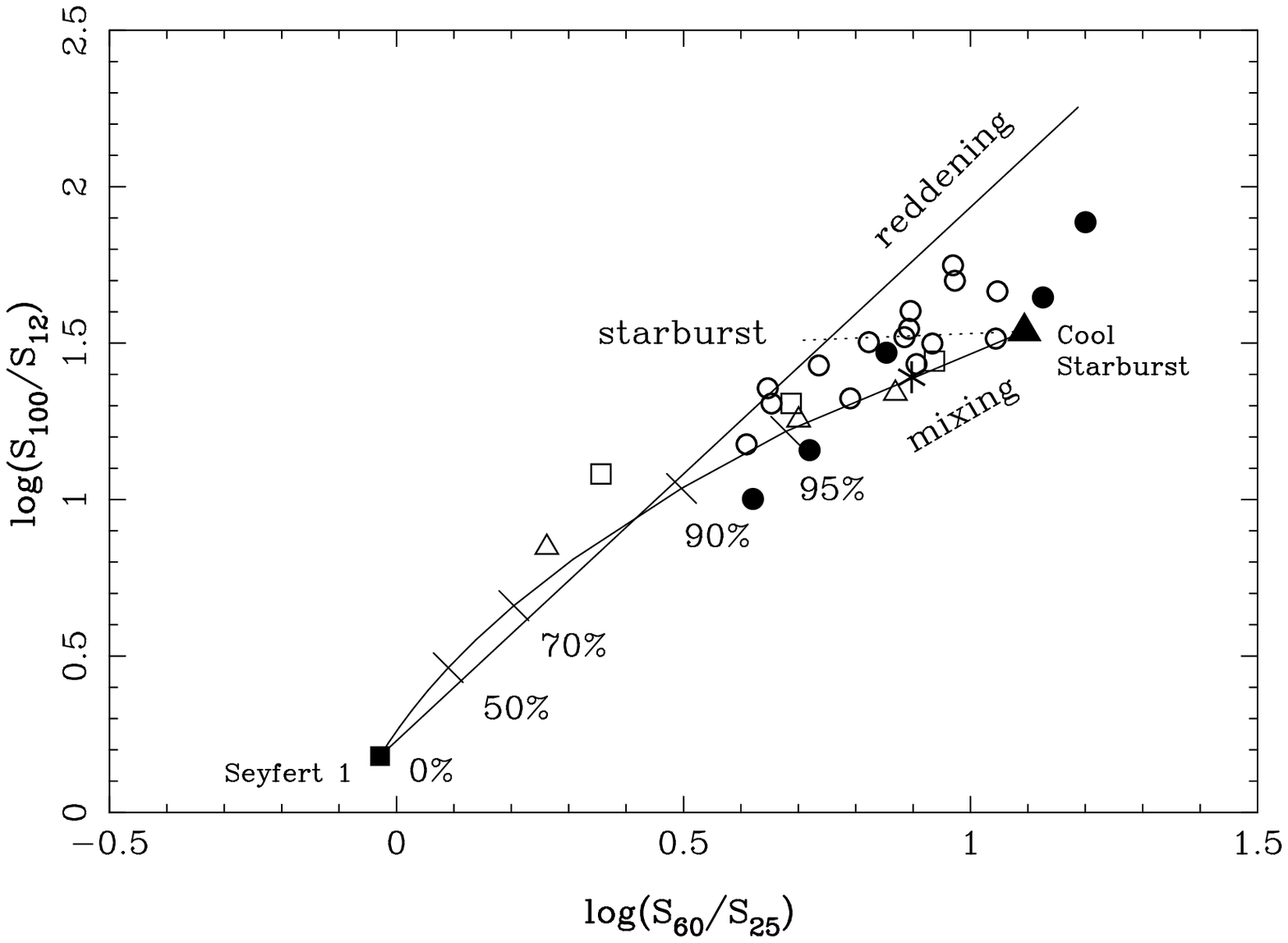} & & \\
\end{tabular}
\caption[hill.fig3a.ps,hill.fig3b.ps,hill.fig3c.ps]
{IR colour-colour diagrams of 
(a) log(S$_{60}$/S$_{25}$) vs log(S$_{25}$/S$_{12}$);
(b) log(S$_{60}$/S$_{12}$) vs log(S$_{100}$/S$_{25}$);
(c) log(S$_{100}$/S$_{12}$) vs log(S$_{60}$/S$_{25}$).
The solid line is the reddening line that shows the effect of
extinction on a typical Seyfert 1 (given by $\blacksquare$), the solid
curve is the extreme mixing curve and the dotted line is an empirical
starburst line from a ``cool'' starburst (given by $\blacktriangle$)
to a ``hot'' starburst, following Dopita {\it et~al.} (1998).
The composite galaxies are shown as: 
$\bullet$ -- detected by PTI; 
$\circ$ -- not detected by PTI; 
$\ast$ -- not observed with PTI. 
The reference galaxies from Table~1 are shown as:
$\triangle$ -- starbursts and $\Box$ -- Seyferts.}
\end{figure}
\clearpage 

\begin{figure} 
\epsscale{1.0}
\figurenum{4}
\plotone{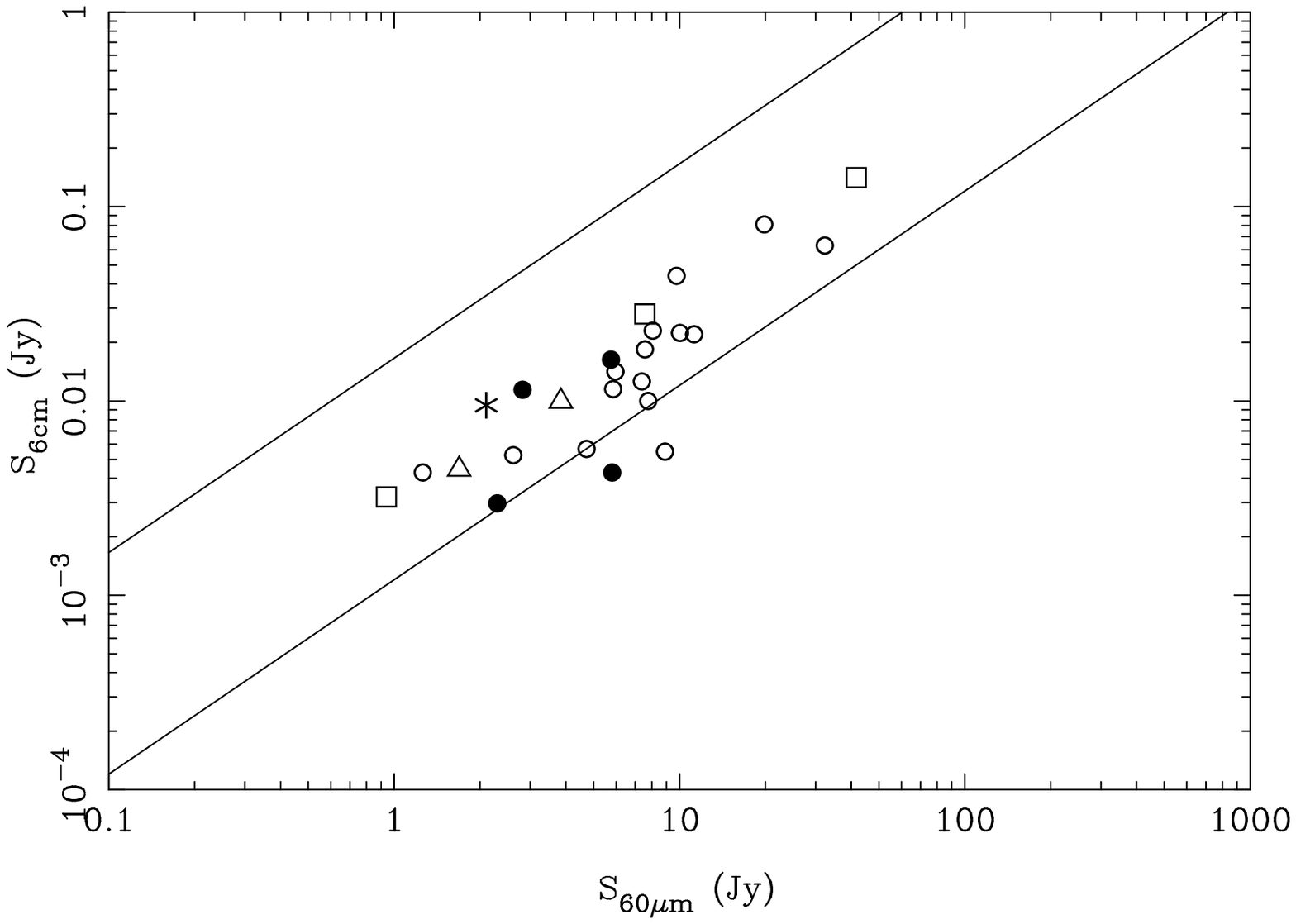}
\caption{Radio-FIR correlation for our sample. The composite galaxies 
are shown as:
$\bullet$ -- detected by PTI; 
$\circ$ -- not detected by PTI; 
$\ast$ -- not observed with PTI. 
The reference galaxies from Table~1 are shown as:
$\triangle$ -- starbursts and $\Box$ -- Seyferts.
The solid lines indicate the $\pm 3\sigma$ extent of the correlation 
found by de~Jong (1985).}
\end{figure}
\clearpage

\begin{figure}
\figurenum{5}
\begin{tabular} {ccc}
\epsscale{0.5}
\plotone{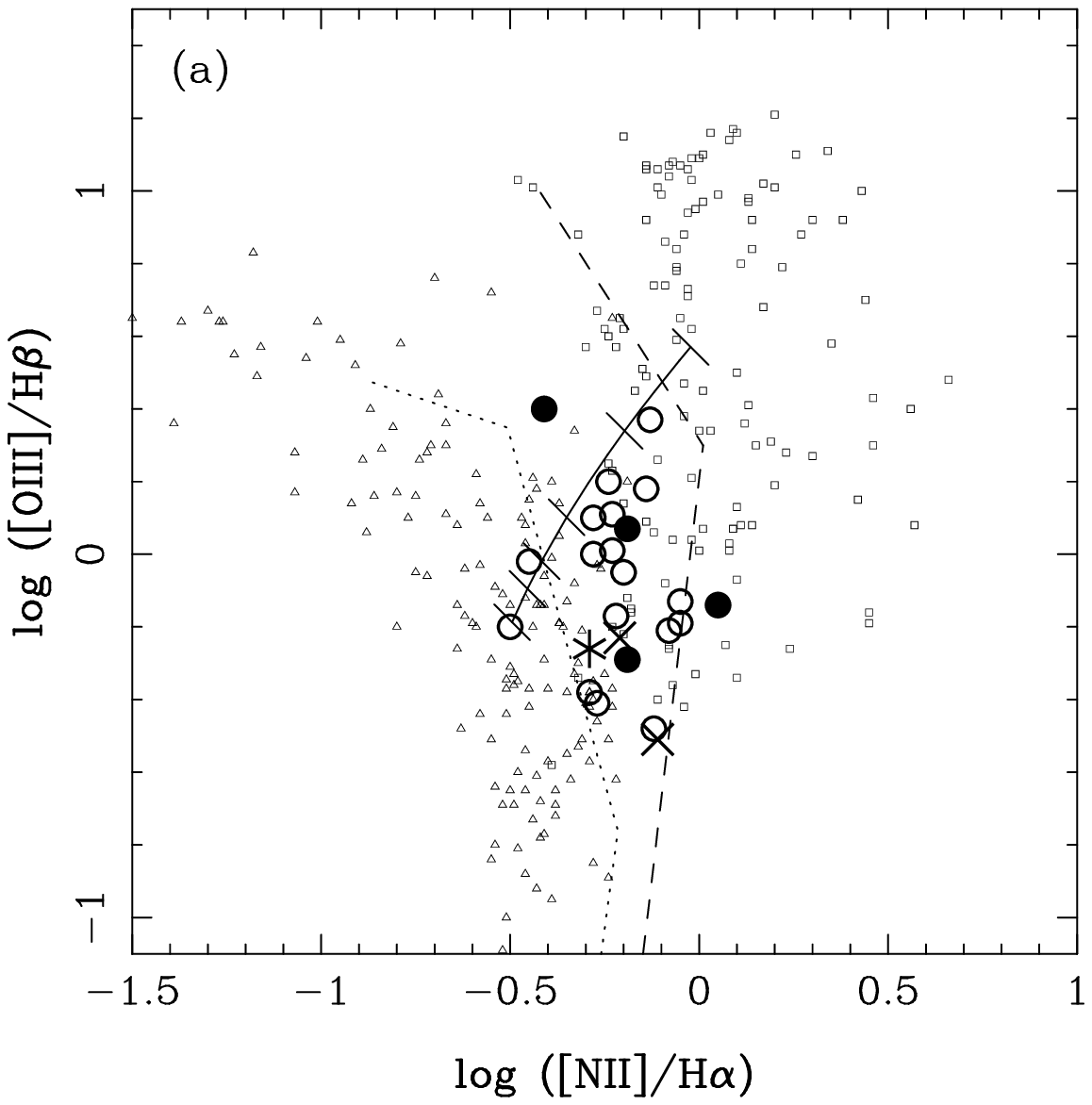} & \plotone{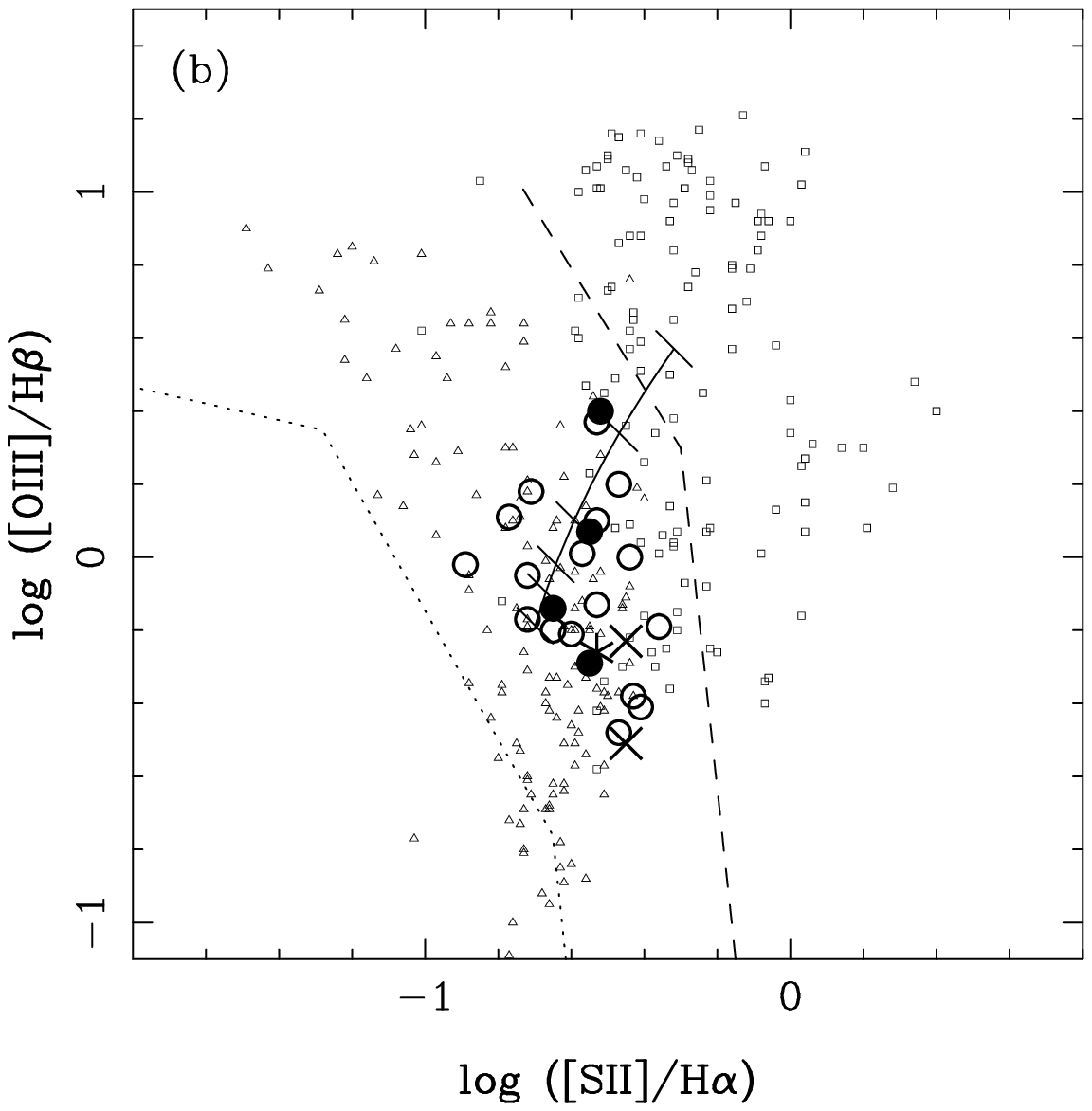} & \\
\plotone{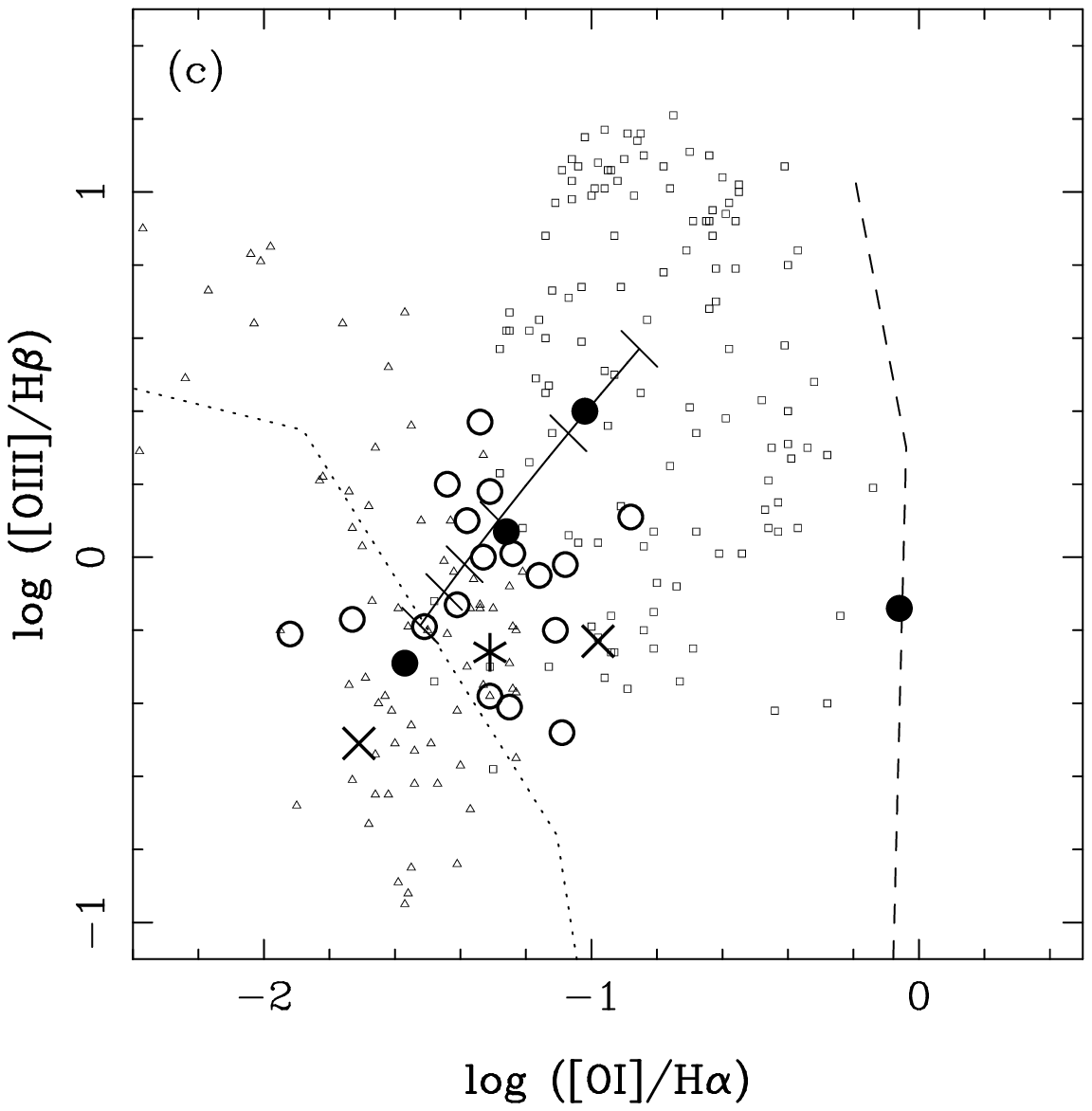} & & \\
\end{tabular}
\caption{The optical diagnostic diagrams:
(a) [O~III]/H$\beta$ vs [N~II]/H$\alpha$;
(b) [O~III]/H$\beta$ vs [S~II]/H$\alpha$;
(c) [O~III]/H$\beta$ vs [O~I]/H$\alpha$.
Symbols have the following meaning: $\triangle$ -- starbursts and
$\Box$ -- AGNs from the literature
(Veilleux \& Osterbrock 1987 and references therein). 
The composite galaxies are shown as: 
$\bullet$ -- detected by PTI; 
$\circ$ -- not detected by PTI; 
$\times$ -- the two individual
galaxies in the galaxy pair of ESO~550-IG025 (a compact core was 
detected with PTI 
directly between the two galaxies); 
$\ast$ -- not observed with PTI. 
The dotted line is a starburst model and the dashed line is a
power-law model from Paper 1. The solid curve is a mixing line with
the tick marks from bottom to top, representing an AGN/starburst mix
of 0\%, 5\%, 10\%, 20\%, 50\% and 100\% AGN. The end-points of the
mixing curve are medians of the data for starburst and Seyfert
galaxies.}
\end{figure}
\clearpage

\begin{figure} 
\figurenum{6}
\epsscale{1.0}
\plotone{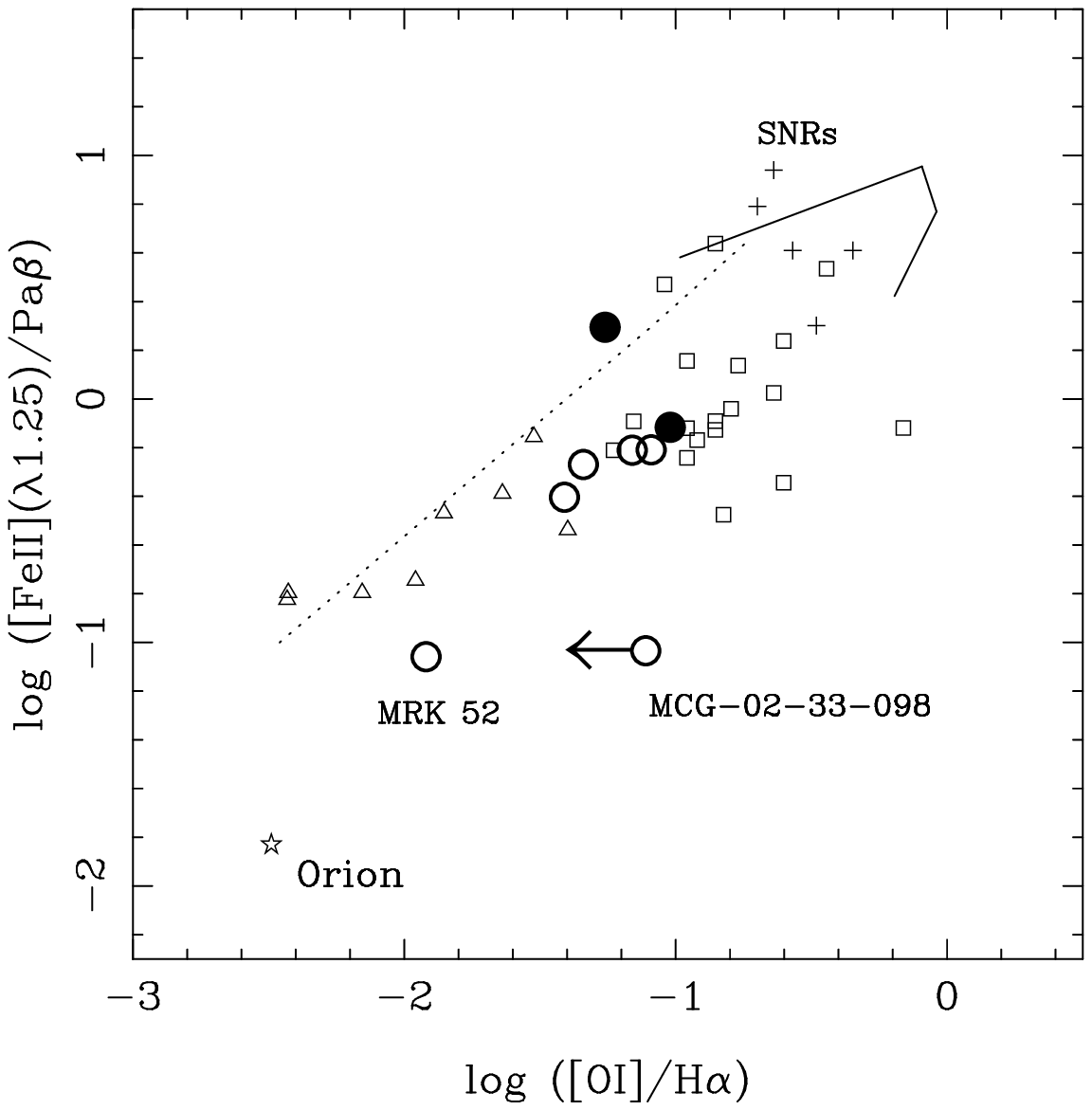} 
\caption{The [Fe~II](1.25 $\mu$m)/Pa$\beta$ vs [O~I]/H$\alpha$ diagram with
the starburst model (dashed line) and power-law model (solid line) from
Paper 1. Symbols have the following meanings:
$\Box$ -- AGNs (Simpson {\it et~al.}\ 1996);
$\triangle$ -- starbursts (Mouri {\it et~al.}\ 1990); + -- SNRs
(Mouri {\it et~al.}\ 1993). The Orion Nebula is
also marked on the diagram (Mouri {\it et~al.}\ 1993). 
The composite galaxies are shown as: 
$\bullet$ -- detected by PTI (Mrk 1344 and MCG+00-29-23); 
$\circ$ -- not detected by PTI.
The optical line ratios are from the surveys of Veilleux \& Osterbrock
(1987), van den Broek {\it et~al.}\ (1991), Ashby {\it et~al.}\ (1995)
and Veilleux {\it et~al.}\ (1995).}
\end{figure}

\end{document}